\documentclass[review]{elsarticle}

\usepackage{lineno,hyperref}
\usepackage{color}
\usepackage{graphicx}
\usepackage{subcaption}
\usepackage{caption}
\usepackage{subcaption}
\usepackage{dcolumn}
\usepackage{amsmath}
\usepackage{amsfonts}
\usepackage{fullpage}
\usepackage{bm}
\usepackage{algorithm}
\usepackage{algorithmicx,algpseudocode}
\usepackage{miller}
\usepackage[bottom]{footmisc}
\usepackage{multirow}
\modulolinenumbers[5]

\usepackage{color}

\journal{Computer Physics Communications}

\graphicspath{{fig/},{fig/vacancy}}

\begin{document}

\begin{frontmatter}

\title{MXE: A package for simulating long-term diffusive mass transport phenomena in nanoscale systems}

\author{Juan Pedro Mendez}
\address{Division of Engineering and Applied Science, California Institute of Technology, Pasadena, CA 91125, USA}

\author{Mauricio Ponga$^*$}
\address{Department of Mechanical Engineering, University of British Columbia, 2054 - 6250 Applied Science Lane, Vancouver, BC, V6T 1Z4, Canada}
\cortext[mycorrespondingauthor]{Corresponding author}
\ead{mponga@mech.ubc.ca}

\begin{abstract}
We present a package to simulate long-term diffusive mass transport in  systems with atomic scale resolution. The implemented framework is based on a non-equilibrium statistical thermo-chemo-mechanical formulation of atomic systems where effective transport rates are computed by using kinematic diffusion laws. Our implementation is built as an add-on to the Large-scale Atomic/Molecular Massively Parallel Simulator (LAMMPS) code. It is compatible with other LAMMPS' functionalities, and shows a good parallel scalability and efficiency. In applications involving mass transport, our framework is able to simulate problems of technological interest for exceedingly large time scales using an atomistic description, which are not reachable with the \emph{state-of-the-art} molecular dynamics techniques. To validate the implementation, we investigated vacancy diffusion, vacancy assisted dislocation climb in metals at high-temperatures, segregation of solutes in free surfaces, diffusion of solutes to grain boundaries, and Hydrogen  diffusion in Palladium nanowires. These examples were validated against known theories, methodologies or experimental results when possible, showing good agreement in all cases. 
\end{abstract}


\end{frontmatter}


{\bf PROGRAM SUMMARY}

\begin{small}
\noindent
{\em Program Title:} MXE package                    \\
{\em Licensing provisions:} GNU GPLv3  \\ 
{\em Programming language:} c++                           \\
{\em Supplementary material:} The user manual and examples are provided along with the source code.                               \\
{\em Nature of problem:} Simulation of mass transport in atomistic systems, involving vacancies, interstitials and solute atoms, is often challenging due to the exceedingly large time scale involved in these processes. Thus, molecular dynamics, a preferred technique to simulate atomistic systems, is not capable of simulating these long-term diffusive phenomena. \\
{\em Solution method:} We present an implementation for simulating mass transport in atomic system. The methodology is based on two main pillars: i) A non-equilibrium thermodynamic formulation of atomic system \cite{ref1}; and ii) Fokker-Planck master equation that encompasses the time evolution of the atomic molar fraction field in atomic systems \cite{ref3}. The proposed implementation is built as an user package of the popular Large-scale Atomic/Molecular Massively Parallel Simulator (LAMMPS). The implementation is flexible and robust, shows good parallel scalability and efficiency, and is compatible with all features available in LAMMPS.\\
{\em Additional comments including Restrictions and Unusual features:} Unique features of the implementation involve the simulation of vacancy, interstitial, solutes and substitutional alloys for exceedingly large time scales. \\
{\em Computer:} Any system with C/C++ compiler. \\
{\em Operating system:} Linux.  \\ 
{\em External routines: MPI, LAMMPS, 12 December 2018 (http://lammps.sandia.gov/)}\\
{\em Restrictions:} The mass transport implementation in this package is limited to the Embedded Atom Model potentials. \\

\end{small}

\newpage

\section{Introduction}\label{Sec:Introduction}


Diffusion of vacancies and solutes in materials is an important phenomena with implications in many technological applications. For instance, in binary alloys, diffusion of the constituents is driven by vacancies that hop from one atomic site to another, promoting the mixture of the elements \cite{Bardeen:1949,Elcock:1958,Manning:1967}. Similarly, in superalloys, vacancies play a crucial role assisting the climb motion of dislocations upon reaching a precipitate \cite{Creep:1999,Reed:2008,Climb:2013}. Thus, the creep velocity and many mechanical properties are dominated by slow and diffusive behavior of vacancies. Mass transport also plays a crucial role in energy storage in solid state devices, where solutes diffuse into a host lattice, storing chemical energy that can be used to power devices \cite{NanoscaleEnergy:2005,Pop:2010}. In all these applications, mass transport is dominated  by mechanisms that have place at the atomic scale and span over long period of times. Due to their technological importance, it is desirable to understand the mechanisms that play a role in such phenomena; however, this is proven to be challenging due to the slow diffusive nature of the problem. Experimental techniques such as atom probe tomography (ATP) can provide useful details in the composition of alloys, specially near defects such as grain boundaries and dislocations offering unique opportunities to understand the effect of alloys in materials \cite{Kelly:2007,Miller:2007,Kelly:2012}. Unfortunately, available experimental techniques cannot offer insights on the mechanisms of diffusion of solutes near defects, that are important to understand and control materials properties in complex environments.

In light of these limitations, modeling techniques can provide useful insights into the mechanisms that play a role in the diffusion of elements, and thus in many mechanical properties of materials. Unfortunately, molecular dynamics (MD) simulations, the preferred tool to understand mechanisms at the nanoscale, has also critical restrictions to study diffusive phenomena. MD simulations 
suffer well known limitations to the size of the time steps that can be used, resulting in simulations with very short windows of time. This makes MD simulations inefficient for capturing long-term diffusive phenomena relevant in mass transport. This hinders usage of MD techniques to simulate and understand diffusive phenomena relevant in mass transport applications. 

Several techniques have been developed in order to extend the time scale accessible by MD, including the HyperMD \cite{HyperMD}, Temperature Accelerated Dynamics \cite{TAD}, and the replica approach \cite{Perez:2009}, to mention but a few. While these techniques have been very useful and have been applied to a variety of applications, they also suffer from limitations in the number of atoms that can be modeled, and are also constrained by the number of rare events that can be tracked during the simulations. For instance, while it is theoretically feasible to simulate diffusion in substitutional and interstitial alloys, the task is challenging, at best, due to the large number of diffusion events that might occur at the atomic level. In order to address these challenges, several methodologies have been developed, including hybrid  Monte Carlo (MC) techniques \cite{MC:2012,MC:2012b}, and phase field approaches \cite{Mianroodi:2019}, that have been successful in simulating segregation of solutes near dislocation cores and defects. Furthermore, coupled diffusive techniques using MD -called diffusive MD (DMD)- have also been recently developed \cite{DMD-Li:2011,Venturini:2014,Romero:2016,Sun:2017,Mendez:2018}, which are outstanding for simulating heat and mass transport at the nanoscale for very large time scale. DMD-like techniques have been shown successful in simulating a wide range of diffusive phenomena in atomistic systems, including sintering \cite{DMD-Li:2011}, Hydrogen (H) diffusion in Palladium (Pd) nanoparticles \cite{Sun:2017}, Lithium (Li) diffusion in Silicon (Si) nanowires \cite{Mendez:2018}, etc. While these works are good examples of mass transport at the atomic scale, DMD implementations are rather difficult, as these formulations require the evaluation of phase field interatomic forces provided by interatomic potentials, which are difficult to compute.
It is then desirable to have such framework implemented in open-source MD codes, as they provide the main computational framework to perform mass transport at the nanoscale. Unfortunately, this is not a trivial task, and previous implementations have been based on specific scripts that modified MD codes \cite{DMD-Li:2011}, or in specific, in-house codes restricted to targeted applications \cite{Romero:2016,Sun:2017,Mendez:2018}, ultimately hindering widespread usage of such techniques.

In this work, we have implemented the DMD method in the LAMMPS code \cite{LAMMPS} to simulate mass transport in atomistic systems. The implementation uses a recently developed model for simulating transport phenomena in discrete systems based on a Fokker-Planck equation \cite{Ponga:2018}. This new approach was inspired on the kinematic mean field theory \cite{Martin:1990}, and is analogous to transport laws used for mass transport in other methods \cite{CurtinDiff:2008,DMD-Li:2011,Mendez:2018}. The proposed kinematic law for transport is then coupled to a non-equilibrium statistical mechanics formulation for atomic systems \cite{Kulkarni:2008,Ariza:2012,Venturini:2014}, allowing the coupled thermo-chemo-mechanical simulation of heat \cite{Ponga:2012,Ponga:2015,Ponga:2016,Ponga:2017} and mass transport at the nanoscale \cite{Romero:2016,Sun:2017,Mendez:2018}. We will refer to the framework (and code) as MXE, short for \emph{maximum-entropy principle}, which is one of the fundamental pillars of the formulation. 

We start the manuscript by briefly reviewing  the theoretical basis of the method in Section~\ref{Sec:Methodology}. Next, we introduce the kinematic transport law for mass transport to encompass the time evolution of the system in Section~\ref{Sec:model}. We discuss the implementation of the framework in the LAMMPS code in Section~\ref{Sec:Implementation}. Then, we exhibit in Section~\ref{Sec:Applications} several applications of mass transport to show the flexibility and capability of the framework and its validation. We study multiple problems of technological relevance involving mass transport in several materials and compare to known models/methodologies and/or experiments. In Section~\ref{Sec:Performance}, we include a detailed study of the computational performance of the package in a high-performance computing facility. Finally, the main conclusions are included in Section~\ref{Sec:Conclusions}.

\section{Background methodology}\label{Sec:Methodology}

As the implementation of this framework is built on previous works \cite{Kulkarni:2008,Ariza:2012,Venturini:2014}, we will quickly review the formulation following Venturini \emph{et al.} \cite{Venturini:2014}. The underlying key of this framework is the Jayne's principle of maximum entropy, which we will referred to as \textbf{mxe} for short. The mxe approach eliminates the need of resolving thermal vibrations of atoms by obtaining thermodynamic forces which drive the kinematic evolution of the system.

Consider a system of $N$ atoms, henceforth referred to as a \emph{site}, and $M$ possible species. For each site $i$, with $i = 1,\ldots, N$, and specie $k$, with $k = 1,\ldots, M$, the following occupancy function is introduced
\begin{equation} \label{Eq:occupancy}
n_{ik} = \begin{cases}  
1, &\ \text{if site $i$ is occupied by specie $k$ }\\
0, &\ \text{otherwise},
\end{cases}
\end{equation}
where the occupation vector $n_{i}$ belongs to the set $\mathcal{O}_{M} = \lbrace \mathbf{n}_i \in \lbrace 0,1\rbrace^M \rbrace$, with the property that every site is always occupied
\begin{equation} \label{Eq:occupancy_relation}
\sum_{i =1}^M n_{ik} = 1.
\end{equation}

The \emph{microscopic state} of the system is characterized by instantaneous positions $\{\mathbf{q}\}=(\mathbf{q}_i)_{i=1}^N \in \mathbb{R}^{3N}$, instantaneous momenta $\{\mathbf{p}\}=(\mathbf{p}_i)_{i=1}^N \in \mathbb{R}^{3N}$ and occupancy array $\{\mathbf{n}\}=(\mathbf{n}_i)_{i=1}^N \in \mathcal{O}_{MN} = (\mathcal{O}_M)^N$. Denoting the probability density function over the phase space as $p(\{\mathbf{q}\},\{\mathbf{p}\},\{\mathbf{n}\}) $, the \emph{expected} or \emph{macroscopic value} of any observable quantity $A(\{\mathbf{q}\},\{\mathbf{p}\},\{\mathbf{n}\})$ is obtained using the well-known phase average in classical statistical mechanics \cite{Landau:2013}
\begin{equation}\label{Eq:phase_average}
\langle A \rangle
= \sum_{ \{\mathbf{n}\} \in \mathcal{O}_{NM} }
\frac{1}{h^{3N}}\int_{\Gamma} A(\{\mathbf{q}\},\{\mathbf{p}\}, \{\mathbf{n}\}))p(\{\mathbf{q}\},\{\mathbf{p}\}, \{\mathbf{n}\}) d\mathbf{q} d\mathbf{p}, 
\end{equation}
where $\Gamma = (\mathbb{R}^{3N} \times \mathbb{R}^{3N})$ represents the phase space, $h$ is the Planck's constant, $h^{3N}$ corresponds to the natural unit of phase volume for systems of \emph{distinguishable particles} and the infinitesimal element of the phase space volume is expressed as
\begin{equation}\label{Eq:phasevolume}
d\mathbf{q} d\mathbf{p}
=
\prod\limits_{i=1}^N
\prod\limits_{j=1}^3
dq_{ij}
dp_{ij}.
\end{equation}
By definition, the expected value of the atomic position and momentum corresponds to the mean position and mean momentum
\begin{subequations}
\begin{align}
& \langle \textbf{q}_i \rangle = \overline{\textbf{q}}_i \\
& \langle \textbf{p}_i \rangle = \overline{\textbf{p}}_i.
\end{align}
\end{subequations} 

As mentioned above, the key assumption of the current framework is the maximization of the information entropy
\begin{equation} \label{Eq:Entropy}
\max_{p} S = -k_B \langle \log{p} \rangle,
\end{equation}
subject to two \emph{local} constraints at \emph{each site}, i.e., 
\begin{subequations} \label{Eq:local_constraints}
\begin{align}
& \langle {n}_{ik} \rangle = x_{ik} \\
& \langle h_i \rangle = e_i,
\end{align}
\end{subequations}
where $k_B$ is the  Boltzmann constant, $x_{ik}$ is the atomic molar fraction of the specie $k$ at site $i$ and $e_i$ is the local energy at site $i$. The maximization of Equation~\ref{Eq:Entropy} subject to the local constraints results in a probability density function of the form
\begin{equation} \label{Eq:probability_density_function}
p = \frac{1}{Z} \exp^{-{ \{ \beta \}^T \{ h \} + \lbrace {\boldsymbol \gamma} \rbrace^T \{ {\bf n} \} } }
\end{equation}
with
\begin{equation} \label{Eq:partition_function}
Z = \sum_{ \{\mathbf{n}\} \in \mathcal{O}_{NM} }
 \frac{1}{h^{3N}} \int_{\Gamma} \exp^{-{\lbrace \beta \rbrace}^T {\lbrace h  \rbrace} + \lbrace {\boldsymbol \gamma} \rbrace^T \lbrace  {\bf n}  \rbrace } d\mathbf{q} d\mathbf{p},
\end{equation}
where $p$ and $Z$ can be regarded as the grand-canonical distribution and grand-canonical partition function of the system, respectively. The quantities $\{ \beta \}$ and $\{ \gamma \}$ are related to the physical quantities of atomic temperatures, $\{ T \}$ and local particle chemical potential, $\{ \boldsymbol \mu \}$, respectively, as 
\begin{equation}
T_i = \frac{1}{k_B \beta_i}.
\end{equation}
and
\begin{equation}
 {\boldsymbol \mu}_i = \frac{  {\boldsymbol \gamma}_i}{\beta_i}.
\end{equation}

Unlike classical statistical mechanics, the local temperature and chemical potential do \emph{not} need to be uniform; this is particularly true when the system is not in an thermodynamic  equilibrium configuration. Using Equations.~\ref{Eq:Entropy}, \ref{Eq:probability_density_function} and \ref{Eq:partition_function}, the grand-canonical total entropy is
\begin{equation} \label{Eq:PhysicalEntropy}
S =  k_B  \{ \beta \}^T   \lbrace e \rbrace - k_B \lbrace {\boldsymbol \gamma} \rbrace^T \lbrace  {\boldsymbol x} \rbrace + k_B \log Z,
\end{equation}
and, consequently, the grand-canonical free entropy is
\begin{equation} \label{eq:Grand-Canonical-Free-Entropy}
\Phi (\lbrace \beta \rbrace, \lbrace \gamma \rbrace) = k_B \log Z (\lbrace \beta \rbrace, \lbrace \boldsymbol {\gamma} \rbrace).
\end{equation}

\subsection{Meanfield approximation}\label{Sec:Meanfield}

As formulated, the probability density function is highly coupled and impractical to compute. A further approximation is used to make this task tractable and efficient. Venturini \emph{et al.} \cite{Venturini:2014} introduced an approximation for the free entropy of the system as
\begin{equation} \label{Eq:Free-energy-approx}
\Phi (\lbrace \beta \rbrace, \lbrace \gamma \rbrace) \geq \Phi_{MF} (\lbrace \beta \rbrace, \lbrace \gamma \rbrace) = k_B \{ \beta \}^T \{ \langle h - h_0 \rangle_0 \} - k_B \log Z_0,
\end{equation}
where $\Phi$ is the real free entropy of the system, $\Phi_{MF}$ is the approximated free-entropy, $h_{0}$ is a local trial Hamiltonian from a family of parametrized trial Hamiltonians, $p_0$ and $Z_0$ are the grand-canonical distribution function and the grand-canonical partition function for the corresponding trial Hamiltonian, and $\langle \ldots \rangle_0$ indicates that the phase average is against $p_0$. Employing wisely trial Hamiltonians that are local, they can be uncoupled, reducing the complexity of the calculations. By choosing a class of local trial Hamiltonians of the form of \cite{Venturini:2014}
\begin{equation} \label{Eq:localHamiltonian}
 h_{0i} = \frac{1}{2m_i} |{\bf p}_i - \overline{{\bf p}}_i|^2 + \frac{m_i \omega_i^2}{2} | {\bf q}_i - \overline{{\bf q}}_i |^2,
\end{equation}
which are parametrized by the \emph{mean field} parameters $\{ \omega \} = (\omega_i)_{i=1}^N \in \mathbb{R}^N$, the mean field probability density function and partition function are
\begin{equation} \label{Eq:MeanFieldRho}
p_0 = \dfrac{1}{Z_0} \prod_{i=1}^N  \exp^{-\beta_i h_{0i}} \exp^{\boldsymbol{\gamma}_i^T\boldsymbol{n}_i},
\end{equation}
and
\begin{equation}
Z_0 = \displaystyle {\prod_{i=1}^N}  \dfrac{\displaystyle \sum_{k=1}^M \exp^{\gamma_{ik}} }{(\hbar \beta_i \omega_i)^3}.
\end{equation}
Thus, the mean field grand-canonical free-entropy of the system is reduced to
\begin{equation} \label{Eq:Grand-Canonical-Free-Energy}
\Phi_{MF} (\lbrace \beta \rbrace, \lbrace \boldsymbol{x} \rbrace) = k_B \sum_{i=1}^N \left[ \beta_i \langle h_i \rangle_0 -3 +3 \log (\hbar \beta_i \omega_i) + \sum_{k=1}^M x_{ik} \log x_{ik} \right].
\end{equation}
The term $\langle h_i \rangle_0$ in Eq.~\ref{Eq:Grand-Canonical-Free-Energy} involves the phase averages of the interatomic potential with respect to the mean-field probability density function $p_0$
\begin{equation}
\langle h_i \rangle_0 = \dfrac{3}{2 \beta_i} + \dfrac{|{\overline{\bf p}_i}|^2}{2m_i} + \langle V_i \rangle_0.
\end{equation}
Sometimes, we refer to $\langle V \rangle_0$ as the \emph{thermalization of the potential} since atomic interactions are computed by taking into account the value of the vibrations and positions at certain temperature. The phase average of the potential includes the effect of the temperature. In our work, we have implemented the corresponding modifications in the code to perform phase averages on the fly by means of third-order Gauss-Hermite quadrature rules. Further discussion of the evaluation of the phase averages can be found in the \ref{Appendix:Sec:Potentials}. Moreover, we have thermalized multiple potentials, including Lennard-Jones (LJ) potential, Morse potential, Embedded Atom Method (EAM) potential and Stillinger and Webber (SW) potential. 

\section{Diffusive transport model at nanoscale}\label{Sec:model}

The previous expression of the grand-canonical free-entropy, Equation~\ref{Eq:Grand-Canonical-Free-Energy}, obtained under the meanfield approximation, requires transport laws to encompass the time evolution of the atomic fields, i.e., temperature, $T_i$, and atomic molar fraction, $x_{ik}$. Now, we will introduce the diffusive transport model that will be used as the backbone of our work. Following Ponga and Sun \citep{Ponga:2018}, we will work towards a master equation that can be applied in both heat and mass transport. This model will only include atomic level information to predict the effective transport rate between sites. 

We consider a system of $N$ particles occupying discretized sites. We place no further limitations on the nature of these sites, so that our model can be used to represent materials with both a lattice structure and those with no structure at all. We also assume some kind of a probability density function for each $i-$th site, $f_i \in [0,1]$, to describe the atomic field. Then, we propose  a master equation that governs the evolution of the probability density $f_i$ as
\begin{equation}\label{Eq:Transport}
\dfrac{\partial f_i(t)}{\partial t} + \tau_r \dfrac{\partial^2 f_i(t)}{\partial t^2} =\sum_{\substack{j=1\\ j\ne i}}^{N}\zeta_{ij}\{[f_j(t)(1-f_i(t))\Gamma_{j\to i}(t)]-[f_i(t)(1-f_j(t))\Gamma_{i\to j}(t)] \},
\end{equation}
where $\zeta_{ij}$ is a pair-wise exchange rate per unit of time between neighboring sites $i$ and $j$, $\Gamma_{i \to j}$ describes the probability of the quantity $f_i(t)$ to be transported from site $i$ to site $j$, and $\tau_r$ the a mean-free time introduced to model ballistic effects, which represents the time for a carrier to be scattered.

We can highlight a few remarks about the master equation for transport. Firstly, the quantity $\Gamma_{i \to j}$ makes no restrictions about the relative energetic levels of the different sites, resulting in a possibility that a quantity could jump from a low-energy state to an high-energy state, although this probability could be very low. Secondly, the summation is carried over all the nearest neighbors of the $i$-th site, which makes our model \emph{local}. However, it is possible to involve more neighbors and make the model non-local, introducing a length-scale in the problem. Although this path is not pursued in this work, the length-scale can be linked to transport properties that are length-dependant in systems. Next, we focus our implementation to perform mass transport in atomic systems. For extension to electronic heat transport in atomistic system, the reader is referred to the recent work of Ullah \emph{et al.} \cite{Ullah:2019}.

\subsection{Mass transport}\label{Sec:Model:Mass}

We next apply the master equation to model the mass transport. We first make the master equation purely diffusive by taking $\tau_r = 0$. By analogy, we interpret the probability density $f_i$ in Equation \ref{Eq:Transport} as the atomic molar fraction $x_i$ for the $i$-th site. The mass transport rate $\Gamma_{i \to j}$ is giving by the Arrhenius equation
\begin{equation}
\Gamma_{i \to j}=\exp(\beta_j \mu_{ij}),
\end{equation} 
where the driven force, or activation energy for the process, is giving by the difference of the chemical potential $\mu_{ij}=\mu_j-\mu_i$ between sites $i$ and $j$. Therefore, the governing equation for the mass transport results as
\begin{equation}\label{Eq:MassTransport}
\frac{\partial  x_{i}}{\partial t}  =  \sum_{{{\substack{ j=1 \\ j \neq i}}}}^{N} D_{ij}[ x_{j} (1-x_{i}) \exp( \beta_j \mu_{ji}  ) - x_{i} (1-x_{j}) \exp( \beta_i \mu_{ij} ) ],
\end{equation} 
where $D_{ij}$ is the pair-wise mass exchange rate that can be computed as
\begin{equation} \label{Eq:attempt_freq_mass}
D_{ij} = \nu_0 \exp (-\beta_i Q_m ),
\end{equation}
where $\nu_0$ is the hoping frequency, and $Q_m$ is the activation energy for the hop of a quantity $x_i$ from the $i-$th to the $j-$th site. The pair-wise mass exchange rate can be related to the macroscopic atom/vacancy diffusivity ($D$) as
\begin{equation} \label{Eq:attempt_freq_mass}
D =\frac{2dD_{ij}}{Zb^2},
\end{equation}
where $Z$ is the coordination number, $b$ is the magnitude of the Burgers vector, $d$ is the dimension of the problem. We remark that Equation \ref{Eq:MassTransport} reduces to the same governing equation that has been presented in previous works \cite{CurtinDiff:2008,DMD-Li:2011,Mendez:2018}.

\section{Implementation}\label{Sec:Implementation}

The proposed framework has been implemented as an user-package in the popular LAMMPS code \cite{LAMMPS} and named as MXE\footnote{A update copy of the package along with examples will be freely available at \cite{webpage:maxent}} package, following the used nomenclature of the source code. The package has been written in $\texttt{c++}$ and parallelized using the Message Parallel Interface (MPI) library. The MXE package uses all attributes of LAMMPS code and is compatible with other subroutines. The implementation is based on two new atom classes (named as $\texttt{atom\_vec\_atomic\_mxe}$ and $\texttt{atom\_vec\_atomic\_mxe\_mt}$) that includes the information of all mxe variables 
($\{ \overline{\bf q} \}$, $\{ \omega \}$, $\{ \bf x \}$, $\{ \beta \}$), several new $\texttt{fix}$ and $\texttt{compute}$ commands to perform specific operations, and the thermalization of interatomic potentials. Every thermalized potential in the new package includes the phase average of the energy and the respective phase average forces, with respect to the mxe variables ($\{ \overline{\bf q} \}$, $\{ \omega \}$, $\{ \bf x \}$, $\{ \beta \}$).

The LAMMPS code is based on the standard velocity-Verlet integrator method. A pseudo-code of Verlet method in LAMMPS can be found in \cite{LAMMPSDeveloperGuide} and is shown in Algorithm \ref{Algorithm:1}. For ease of visualization, Algorithm \ref{Algorithm:1} only includes the most relevant sections to understand the functioning of the MXE package. The readers are referred to the LAMMPS developer guide \cite{LAMMPSDeveloperGuide} for further description of the internal structure. In LAMMPS, any operation and computation to a group of atoms is performed by $\texttt{fix}$ and $\texttt{compute}$ commands. The $\texttt{fix}$ command is executed every time step during the simulation. Furthermore, this operation can be called at different moments during the execution of the code as shown in Algorithm \ref{Algorithm:1}: before updating or computing the list of neighbours, using $\texttt{fix}\rightarrow\texttt{initial\_integrate()}$ or $\texttt{fix}\rightarrow\texttt{post\_integrate()}$; before computing the forces, using $\texttt{fix}\rightarrow\texttt{pre\_force()}$; after computing the forces, using $\texttt{fix}\rightarrow\texttt{post\_force()}$ or $\texttt{fix}\rightarrow\texttt{final\_integrate()}$ or $\texttt{fix}\rightarrow\texttt{end\_of\_step()}$.
\begin{algorithm}
\caption{A simplified pseudo-code of Verlet method in LAMMPS \cite{LAMMPSDeveloperGuide}. This pseudo-code only includes all the key sections to understand the MXE package.}
\label{Algorithm:1}
\begin{algorithmic}[1]
\State Define and initialize the problem
\Loop ~over N timestep:
\State fix$\rightarrow$initial$\_$integrate(): $\texttt{mass\_transport/atom}$
\State fix$\rightarrow$post$\_$integrate(): $\texttt{setatomicmolar}$
\State Update or compute the neighbour list and establish the communication between the processors
\State Set the forces to zero
\State fix$\rightarrow$pre$\_$force()
\State Compute the forces and establish the communication between the processors
\State fix$\rightarrow$post$\_$force(): $\texttt{setchemicalpotential}$
\State fix$\rightarrow$final$\_$integrate(): $\texttt{optfreq}$
\State fix$\rightarrow$end$\_$of$\_$step()
\State call the defined computes and print the outputs for this timestep
\EndLoop
\end{algorithmic}
\end{algorithm}

In this implementation, we evolve the system with time by a staggered scheme consisting in two main operations at each timestep: i) diffusive step (mass transport), and ii) relaxation of the configuration. During the diffusive step, we fix the atomic positions and the frequencies, whereas the atomic molar fraction is updated by integrating Equation~\ref{Eq:MassTransport}. The integration time step is several orders of magnitude larger than the critical time step in MD. The framework is able to simulate exceedingly large windows of time in problems that are out of the scope of MD techniques. Followed the first step, the system is then relaxed by maximizing the grand-canonical free-entropy (Equation \ref{Eq:Grand-Canonical-Free-Energy}), in which the atomic positions and frequencies are updated, while the atomic molar fraction is fixed. These two operations are integrated in the LAMMPS code using $\texttt{fix}$ commands.

The diffusive step for the mass transport is carried out by the new $\texttt{fix}$ $\texttt{mass\_transport/atom}$ fix command. This $\texttt{fix}$ is called at very beginning of each timestep using $\texttt{fix}\rightarrow\texttt{initial\_integrate()}$. The command update the atomic molar fractions $\{ x\}$ at every time step following the governing master Equation~\ref{Eq:MassTransport}. The integration is carried out using an Euler's forward integration scheme. Therefore, a special care must be taken as the stability of the integration scheme depends strongly upon the choice of the integration time step.

For the equilibrium step, we use dynamic relaxation method by using a Verlet integration scheme with a damping factor for both, positions and frequencies. Alternatively, the macroscopic positions can be updated at every time step using the $\texttt{fix}$ $\texttt{NVE}$, or $\texttt{NPT}$ fix commands, setting the temperature to $\sim 0K$, or using any energy minimization method implemented in LAMMPS. The update of the the atomic frequencies $\{\omega\}$ at every time step is carried out by the new $\texttt{fix}$ $\texttt{optfreq}$ fix command which performs a dynamic relaxation optimization. This command has been implemented using $\texttt{fix}\rightarrow\texttt{final\_integrate()}$, so it is called at very end of each time step of the simulation.

Furthermore, the MXE package includes other new features: $\texttt{fix}$ commands, $\texttt{setchemicalpotential}$ and $\texttt{setatomicfraction}$; the $\texttt{compute}$ $\texttt{fe}$ command; and the $\texttt{delete\_atoms}$ command. The fix $\texttt{setchemicalpotential}$ and $\texttt{setatomicfraction}$ commands are used to set the chemical potential and the atomic molar fraction on each atom in a specified group of atoms to a specified value, respectively. This is useful when one wants to simulate diffusion of species since they generate a chemical gradient. The compute $\texttt{fe}$ commands computes the total free entropy of the system (Equation \ref{eq:Grand-Canonical-Free-Entropy}); and, the $\texttt{delete\_atoms}$ command of the package has been updated to delete atoms whose atomic molar fraction is smaller than a specified value in a specified group of atoms. The calls of the new fix commands during the Verlet algorithm are highlighted in Algorithm~\ref{Algorithm:1}. The commands are explained in the user manual provided with the package. 

\section{Mass transport applications} \label{Sec:Applications}

After introducing the mxe framework and its implementation in LAMMPS, we proceed to present some examples that involve mass transport in atomistic systems. The main goal is to exhibit the capabilities and the flexibility of the MXE package for prospective applications, and validate it against known solutions when possible. The simulation results are visualized using OVITO \cite{Stukowski:2009}. Several test cases were implemented and validated against MD. These involve lattice parameters and stacking fault energies at different temperatures in several materials, etc. These are omitted and shown in the user manual with the corresponding scripts.

\subsection{Vacancy diffusion in Copper (Cu)} \label{Sec:Applications:Vacancy_diffusion}

Diffusion of a single vacancy is modelled in a face centered cubic (fcc) Cu single crystal. We use a simulation cell of dimension of $l_x = 60 a_0 \times l_y = 60 a_0 \times l_z = 60 a_0$, where $a_0 = 0.3615$ nm is the lattice parameter, containing a total of 864,000 sites. We apply periodic boundary conditions in all directions and constant temperature during the whole simulation. This simulation is similar to the work of Li \emph{et al.} \cite{DMD-Li:2011}, however, our computational cell is larger and the concentration of vacancies in the simulation is comparable with the equilibrium concentration. In our simulations, a single vacancy is generated by initializing the atomic molar fraction of the central atom to zero, whereas the remaining sites of the sample are set to one. The total vacancy concentration of the sample is $c_{vac} = 1.15 \times 10^{-6}$, i.e., a vacancy concentration of $9.8 \times 10^{-5}$ vacancies$\cdot$nm$^{-3}$, which is very similar to the equilibrium value. Regarding the diffusive kinetic law, we used the parameters $\nu_0= 10^{13}$ s$^{-1}$ and $Q_m = 0.69$ eV in our simulations, chosen to match the experimental vacancy diffusivity in bulk Cu. Using a NPT ensemble, we reduced the pressure to zero at different temperatures, and let the concentration of vacancies evolve with time in the sample, governed by the kinetic law.

\begin{figure}
\centering
\includegraphics[width=0.3\textwidth]{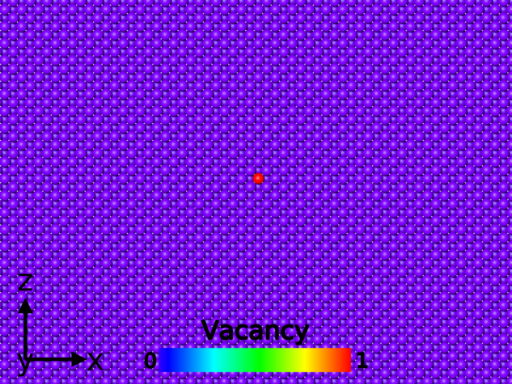}
\includegraphics[width=0.3\textwidth]{vacancy/T100K.png}
\includegraphics[width=0.3\textwidth]{vacancy/T100K.png} \newline
\includegraphics[width=0.3\textwidth]{vacancy/T100K.png}
\includegraphics[width=0.3\textwidth]{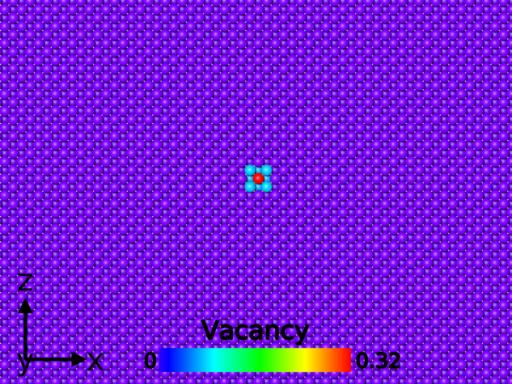}
\includegraphics[width=0.3\textwidth]{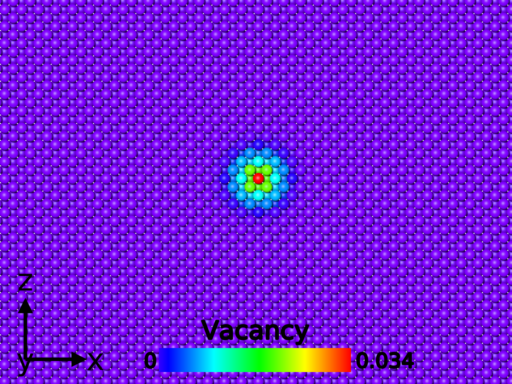} \newline
\caption{The diffusion of a single vacancy in fcc-Cu at $T=100$ K after: (a) $t=0$ ms, (b) $t= 5 \times 10^{-4}$ ms, and (c) $t=3.22$ ms; and at $T=300$ K after (d) $t=0$ ms, (e) $t = 5 \times 10^{-4}$ ms and (f) $t=3.22$ ms}
\label{vacancy-diffusion-1}
\end{figure}

\begin{figure}
\centering
\includegraphics[width=10cm]{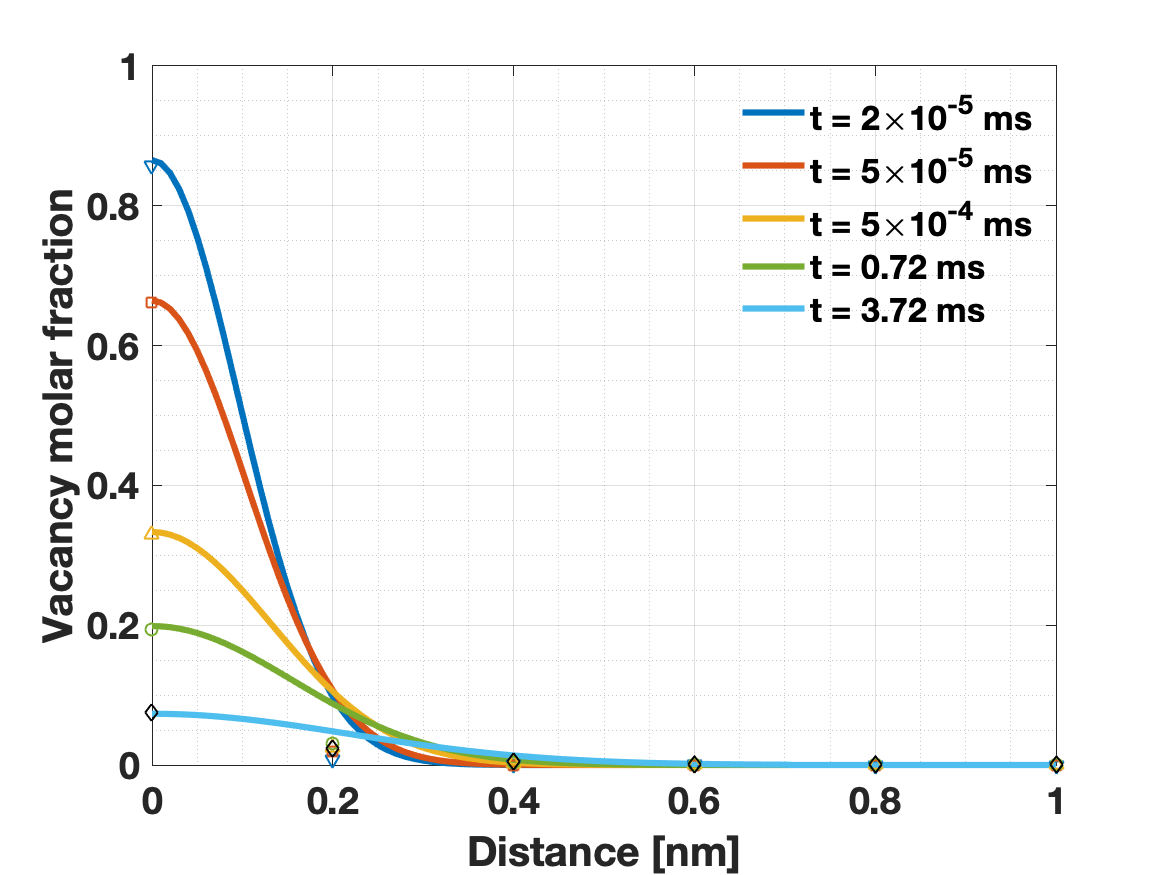}
\caption{Evolution of the vacancy molar fraction at $T = 300$ K at a distance $d$ from the initial position of the vacancy. The result also shows the approximation by the mass diffusion equation for a mass point in an infinite continuum system.}
\label{vacancy-diffusion-2}
\end{figure}

Figure \ref{vacancy-diffusion-1} shows the diffusion of a single vacancy in fcc-Cu at $T=100$ K and $T=300$ K. At low temperature, e.g. $T=100$ K, the vacancy does not have enough entropic effects to overcome the energy barrier to diffuse, so the vacancy stays immobile at the initial position during the simulation. On the other hand, when temperature was increased to $T=300$ K, the vacancy diffused faster as shown in Figure~\ref{vacancy-diffusion-1}. Figure~\ref{vacancy-diffusion-2} represents the evolution of the average atomic molar fraction at a distance $d$ from the original position of the vacancy at $T=300$ K. The obtained result can be approximated by the solution to a diffusion equation for a mass point in an infinite continuum system \cite{crank1979}, $c(d,t) = \Omega^{-1} (4 \pi D t)^{-3/2} e^{-d^2/(4 D t)} $, where $\Omega$ is the atomic density. As expected, the vacancy dilutes radially because of the lack of defects and the applied periodic boundary conditions in all directions. The vacancy does not have any preferences in the direction of diffusion. Once the equilibrium of the system is reached, the vacancy concentration tends to be constant everywhere in the sample; it can be understood as in equilibrium the probability of finding a vacancy in the sample without defects is the same everywhere.

\subsection{Dislocation climb of a edge dislocation in Molybdenum}

We now turn our attention to dislocation climb due to vacancies diffusion towards dislocation cores. This is a complex problem in many applications in materials science and mechanics of materials, that cannot be simulated with MD simulations due to the long time scale of the problem. First, we generated a Molybdenum (Mo) body centered cubic (bcc) perfect crystal oriented along $x: [1 1 1]$, $y: [\overline{1} 0 1]$, $z: [1 \overline{2} 1]$ with dimensions of $47.8$ nm $\times   1.8$ nm $\times 45.6$ nm. We used the embedded atom method potential developed by Zhou \emph{et al.}  \cite{Zhou:2004} to simulate the atomic interactions. A single $\langle 111 \rangle \{112\}$ edge dislocation was generated by removing half plane of atoms, following the work of Gr\'egoire \emph{et al.} \cite{Gregoire:2017}. Periodic boundary conditions were applied along the dislocation line sense ($y-$direction) and traction-free surfaces were taken in $x-$ and $z-$directions. In order to generate an external stress, we applied a small stretching in the sample ($\lambda = 1.0001$) which stretched the atoms by 0.01\% with respect to the equilibrium position of the atoms near the dislocation. 

\begin{figure}
\centering
\includegraphics[width=0.48\textwidth]{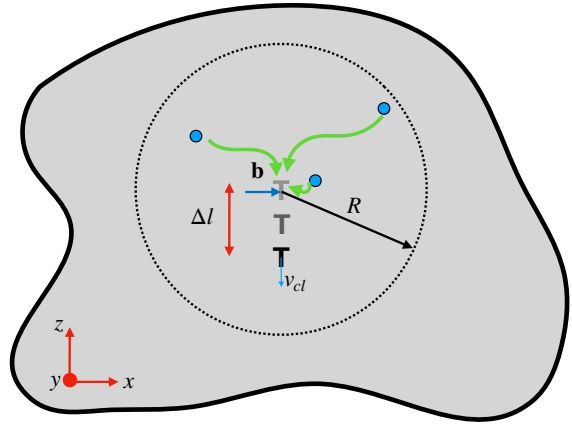}
\caption{Schematic picture of the setup used in the simulation of vacancy assisted climb of an edge dislocation in bcc-Mo. Vacancies (in blue) migrate to the core due to the higher chemical potential near the core. The dislocation line (shown with $\perp$) moves downwards  when vacancies are annihilated. The resulting displacement ($\Delta l$) and the diffusion time allow to compute a climb velocity ($v_{cl}$).}
\label{Fig:ClimbSchematic}
\end{figure}

The temperature of the sample was set constant in the whole simulation to $T=1400$ K, which is approximately half the melting point of Mo. The vacancy formation and migration energy were taken to be $E^v_f = 3.0$ eV and $E^v_m = 1.3$ eV, respectively \cite{Mattsson:2009}, and the attempt frequency to $\nu_0 = 10^{12}$ s$^{-1}$. In order to favour vacancy diffusion to the core, we generated a reservoir of vacancies far from the dislocation core. The reservoir was given by all atoms \emph{outside} a cylinder centered at the dislocation with a radius $R= 30 b$, where $b = 0.272$ nm is the Burgers vector. For these atoms, the chemical potential was set to $\mu \sim -9.7$ eV. This value was always lower than the chemical potential in the core and its surroundings. A schematic of our setup is shown in Figure \ref{Fig:ClimbSchematic}. 

\begin{figure}
\centering
\begin{subfigure}[b]{0.48\textwidth}
\includegraphics[width=\textwidth]{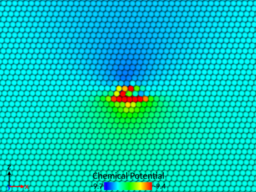}
\caption{}
\end{subfigure}
\begin{subfigure}[b]{0.48\textwidth}
\includegraphics[width=\textwidth]{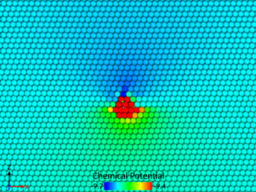}
\caption{}
\end{subfigure}
\caption{Evolution of the chemical potential during the dislocation climb process. We see that the core has a higher chemical potential and attracts vacancies. Values are in eV.}
\label{Fig:Chem}
\end{figure}

A common assumption  in climb of dislocations at high temperature is that the dislocation  line contains a saturation of jogs in equilibrium \cite{Hirth}. In our simulations, we generated a jog by placing a vacancy in the core. The effect of this vacancy was to generate a small jog together with a chemical potential gradient, as can be seen in Figure \ref{Fig:Chem}(a). From the figure, we observe that the dislocation has a higher chemical potential in comparison with the remaining of the simulation cell, which indicates that vacancies will tend to migrate over there. Of remarkable interest, we also see that the half plane with an extra layer of atoms (bottom) has larger chemical potential than the other half plane (up). This is expected, as the atoms in the bottom half part of the simulation are closer to each other (compression) and the atoms in the upper half side are further away to each other (tensile state). Thus, vacancies tend to migrate to the bottom half of the simulation, reducing the chemical potential gradient. As vacancies migrate, the dislocation climbs down by annihilating vacancies. Figure \ref{Fig:Chem}(b) shows the chemical potential for the dislocation after the dislocation has climbed down about 1 nm. We see that the core has migrated, and with it the chemical potential distribution.

\begin{figure}
\centering
\begin{subfigure}[b]{0.32\textwidth}
\includegraphics[width=\textwidth]{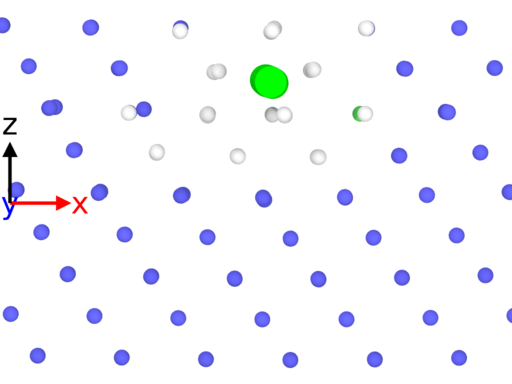}
\caption{}
\end{subfigure}
\begin{subfigure}[b]{0.32\textwidth}
\includegraphics[width=\textwidth]{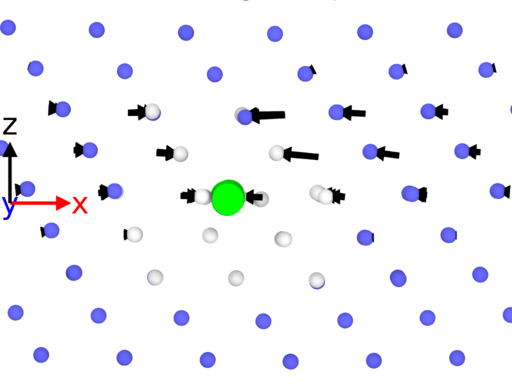}
\caption{}
\end{subfigure}
\begin{subfigure}[b]{0.32\textwidth}
\includegraphics[width=\textwidth]{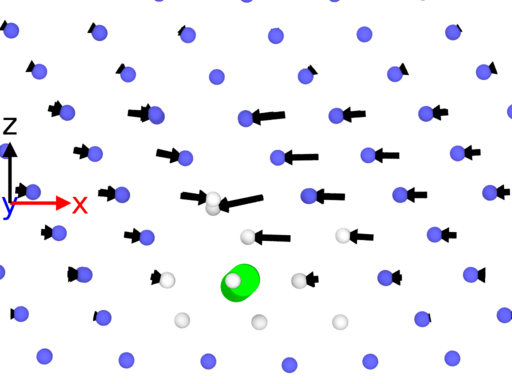}
\caption{}
\end{subfigure}
\caption{Mechanism for edge-dislocation climb in bcc-Mo. Dislocation line is shown in green; atoms are colored according to their local structure: bcc (blue), and other structures (white). (a) Initial dislocation core, (b) Dislocation core after 10 vacancies have been annihilated, and (c) after 18 vacancies have been annihilated. The arrows show the atomic displacement of each atom with respect to the initial position shown in (a). The reference system is fixed in all figures.  }
\label{Fig:Chem2}
\end{figure}

Let us now focus on the atomistic details of the mechanism for vacancy assisted climb of an edge dislocation in bcc-Mo. Figure \ref{Fig:Chem2}(a) shows a front view of the atoms in the initial position. The atoms are colored using the common neighbor analysis (CNA) algorithm using OVITO \cite{Stukowski:2009,CNA}. We see that atoms in the \emph{core} are shown in white, while atoms in bcc structure are shown in blue. The dislocation line, obtained using the dislocation extraction algorithm (DXA) \cite{DXA}, is shown in green. The reference system shown in the picture is at the same location in all the pictures of Figure \ref{Fig:Chem2}. Figure \ref{Fig:Chem2}(b) shows the dislocation core after 10 vacancies have migrated and annihilated. The time scale is computed to be around $t=25000$ s, an overwhelming long time scale for MD simulations. We see that the dislocation line (green line) has moved down $\Delta l = 0.6$ nm. The atomic displacements with respect to the positions in Figure \ref{Fig:Chem2}(a) are shown with the black arrows in in Figures \ref{Fig:Chem2}(b)-(c). As depicted in the Figure, when vacancies are annihilated in the core, the atoms surrounding the dislocation move towards the core, generating a motion of the dislocation downwards. This procedure is repeated as more vacancies are annihilated in the core, as shown in Figure \ref{Fig:Chem2}(c) when 18 vacancies have been annihilated. The total displacement of the dislocation is about $\Delta z = 1.2$ nm and the total simulated time is computed to be around $t = 50000$ s. 

Having computed the total displacement and the elapsed time, one can compute the climb velocity as $v_{cl}^{\text{mxe}}= 2.4 \times 10^{-5}$ nm$\cdot$s$^{-1}$. Continuum models for vacancy assisted climb motion of edge dislocations gives a climb velocity of \cite{Hirth}

\begin{equation} \label{eq:ClimbVelocity}
v_{cl} = \frac{2\pi}{\ln (R/b)}\frac{D}{b}\bigg[\exp{\bigg(\frac{\sigma_{xx}\Omega}{k_BT} \bigg)}-1\bigg].
\end{equation}
where $R = 30 b$ is a distance where the vacancy have reach the equilibrium concentration, $D = 9 \times 10^{-22}$ m$^2\cdot$s$^{-1}$ is the vacancy diffusivity in Mo \cite{Mattsson:2009}, $\sigma_{xx} = 10$ MPa is the stress applied, and $\Omega \approx b^3$ is the volume of the vacancy. Using Equation \ref{eq:ClimbVelocity} we obtain a velocity of $v_{cl} = 6.4 \times 10^{-5}$ nm$\cdot$s$^{-1}$, which is in remarkably agreement with our simulations. The difference between the continuum mechanics law and the direct atomistic simulation is less than one order of magnitude.

\begin{figure}
\centering
\begin{subfigure}[b]{0.45\textwidth}
\includegraphics[width=\textwidth]{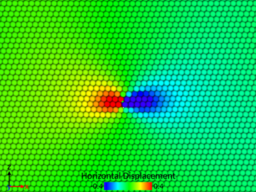}
\caption{}
\end{subfigure}
\begin{subfigure}[b]{0.45\textwidth}
\includegraphics[width=\textwidth]{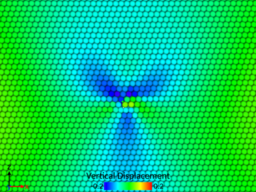}
\caption{}
\end{subfigure}
\caption{Atomic displacement for atoms near the core during the dislocation climb. (a) horizontal displacement, and (b) vertical displacement with respect to the original dislocation configuration.}
\label{Fig:Chem3}
\end{figure}

Finally, to show the complexity of dislocation climb, we illustrate the displacement field with respect to the original dislocation in Figure \ref{Fig:Chem3}. The horizontal displacement (see Figure \ref{Fig:Chem3}(a)) indicates that the atoms surrounding the core are attracted to it, due to the vacancy migration. This atomic motion is also three-dimensional, with some of the atoms displacing in the vertical direction to accommodate the displacement (see Figure \ref{Fig:Chem3}(b)). 

\subsection{Mg segregation on surfaces in Al alloys} \label{Sec:Applications:Surface_segregation}

In order to further benchmark our implementation against known methods, we simulate Mg segregation on diverse free surfaces in Al-4at\%Mg alloys at $T=600$ K to compare against Monte Carlo (MC) simulations performed by Liu \emph{et al.} \cite{Liu:1997}. The empirical potential used to model the Al-Mg alloy was also introduced in the same work \cite{Liu:1997}, and the measured microscopy diffusion coefficient for the kinetic equation to model the Mg diffusion at $T=600$ K was chosen to be $D = 5.4914 \times 10^{-16}$ m$^2 \cdot$s$^{-1}$ \cite{brandes:1992}. For setting up the simulations, we consider an infinite slab in two directions ($y$ and $z$), where we applied periodic boundary conditions, and finite size in the third direction ($x$), where we applied traction-free boundary conditions. The crystallographic plane on the free surfaces corresponded to (100), (110) and (111) in our simulations. Following \cite{Liu:1997}, the length of the sample was set to 40, 50 and 36 layers with 128, 96 and 144 atoms per layer, respectively. Initially, every site of the sample was initialized with 96\% of Al and 4\% of Mg. Then, we allowed the system to reach equilibrium at a constant temperature of $T=600$ K following maximization of the free entropy of the system. 
Note that the mass of the system was preserved during the whole simulation.

\begin{figure}
\centering
\begin{subfigure}[b]{0.32\textwidth}
\includegraphics[width=\textwidth]{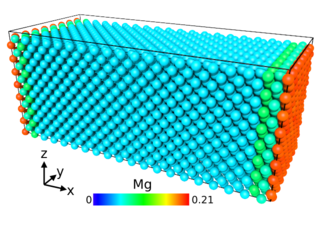}
\caption{}
\end{subfigure}
\begin{subfigure}[b]{0.32\textwidth}
\includegraphics[width=\textwidth]{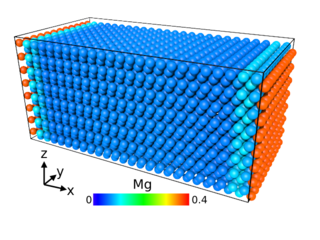}
\caption{}
\end{subfigure}
\begin{subfigure}[b]{0.32\textwidth}
\includegraphics[width=\textwidth]{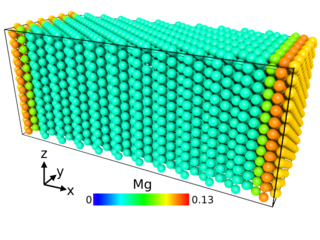}
\caption{}
\end{subfigure}
\caption{Mg segregation on (a) (100) free surface, (b) (110) free surface and (c) (111) free surface in Al-4\%Mg alloys at $T = 600$ K after $t=0.1$ millisecond.}
\label{fig:surface_segregation_1}
\end{figure}

\begin{figure}
\centering
\begin{subfigure}[b]{0.32\textwidth}
\includegraphics[width=\textwidth]{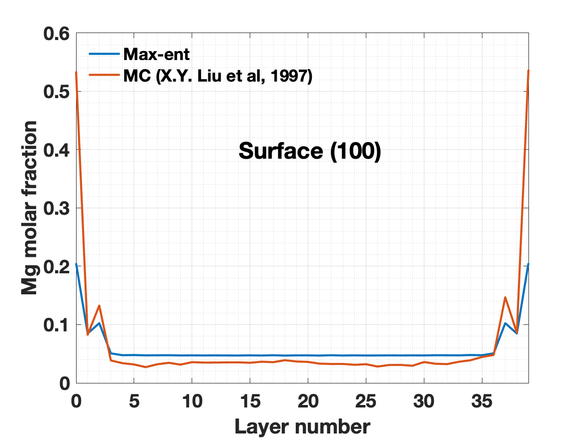}
\caption{}
\end{subfigure}
\begin{subfigure}[b]{0.32\textwidth}
\includegraphics[width=\textwidth]{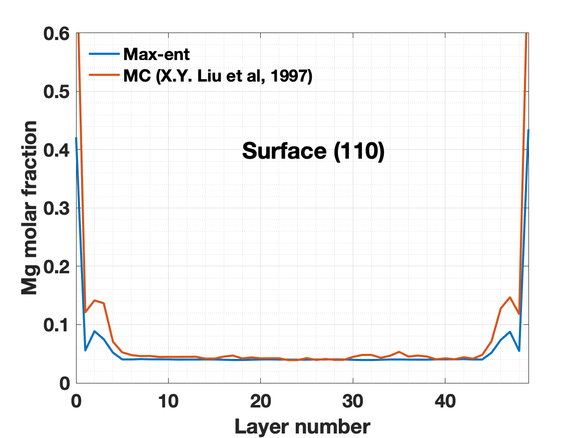}
\caption{}
\end{subfigure}
\begin{subfigure}[b]{0.32\textwidth}
\includegraphics[width=\textwidth]{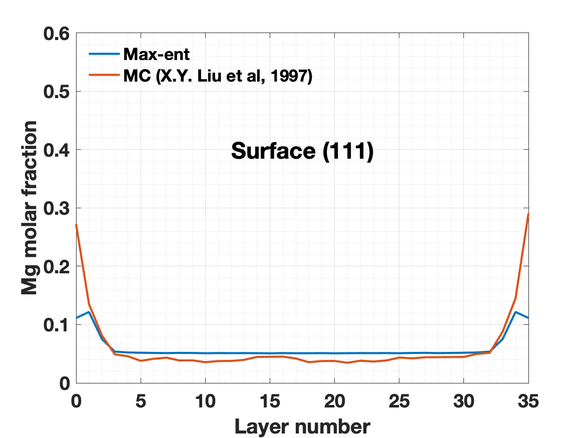}
\caption{}
\end{subfigure}
\caption{Mg atomic molar fraction on different layers for the slabs presented in Fig~\ref{fig:surface_segregation_1} corresponding to the free surface (a) (100), (b) (110) and (c) (111) in Al-4\%Mg alloys at $T = 600$ K.}
\label{fig:surface_segregation_2}
\end{figure}

Figure~\ref{fig:surface_segregation_1} shows the Mg atomic molar fraction of the slab for each site at the equilibrium configuration for the three orientations after $t =0.1$ ms. For all cases, the diffusion of Mg towards the free surface is seen, however, the values of the atomic molar fraction are different for each orientation and layer. Figure~\ref{fig:surface_segregation_2} compares the average of the Mg atomic molar fraction for each layer parallel to the free surface against the MC simulations of Liu \emph{et al.} \cite{Liu:1997}. The results evidence a strong anisotropic behavior, i.e., the Mg segregation depends on the crystallographic orientation. The range of Mg segregation only reaches a few atomic layers from the surface as shown in Figures~\ref{fig:surface_segregation_1} and \ref{fig:surface_segregation_2}. In general, these results are in good agreement with \cite{Liu:1997}, both methods reproduce similar behavior, except for the (111) surface where some discrepancies arise as shown in Figure~\ref{fig:surface_segregation_2}(c).

\subsection{Effect of grain boundaries in Cu-Ni alloys}

\begin{figure}
\centering
\begin{subfigure}[b]{0.45\textwidth}
\includegraphics[width=\textwidth]{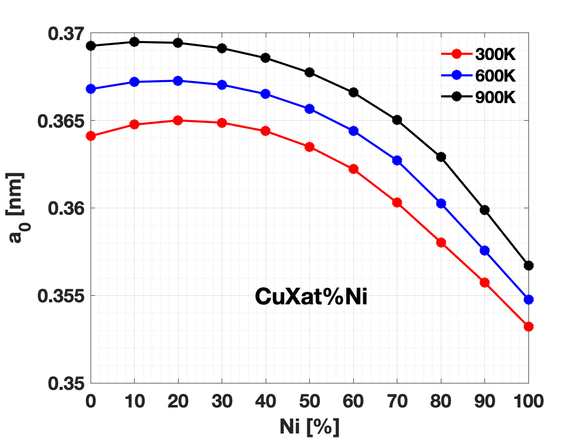}
\caption{}
\end{subfigure}
\begin{subfigure}[b]{0.45\textwidth}
\includegraphics[width=\textwidth]{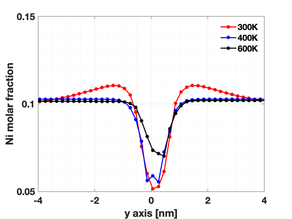}
\caption{}
\end{subfigure}
\caption{(a) Lattice parameter for homogeneous Cu-x at.\% Ni alloys, where x is the atomic molar fraction of Ni, at different temperatures. (b) Average of the Ni atomic molar fraction as a function of the distance from the GB for the results presented in Figure~\ref{Fig:CuNialloy-2}.} 
\label{Fig:CuNialloy-1}
\end{figure}

\begin{figure}
\centering
\begin{subfigure}[b]{0.3\textwidth}
\includegraphics[ width=\textwidth]{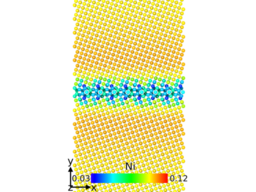}
\caption{}
\end{subfigure}
\begin{subfigure}[b]{0.3\textwidth}
\includegraphics[width=\textwidth]{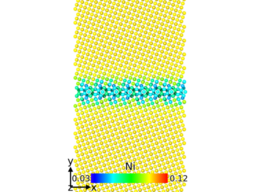}
\caption{}
\end{subfigure}
\begin{subfigure}[b]{0.3\textwidth}
\includegraphics[width=\textwidth]{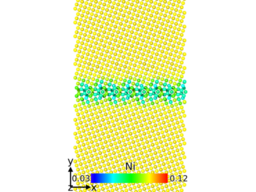}
\caption{}
\end{subfigure}
\caption {Atomic configurations of the atoms near the grain boundary after an annealing process at constant temperature during 8.20 s. Color map shows the Ni atomic molar fraction for an initial Cu-10at\%Ni alloy: (a) $T=300$ K, (b) $T=600$ K and (c) $T=800$ K.} 
\label{Fig:CuNialloy-2}
\end{figure}

\begin{figure}
\centering
\begin{subfigure}[b]{0.24\textwidth}
\includegraphics[width=\textwidth]{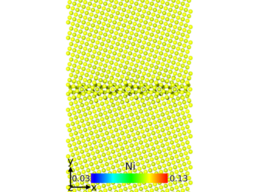}
\caption{}
\end{subfigure}
\begin{subfigure}[b]{0.24\textwidth}
\includegraphics[width=\textwidth]{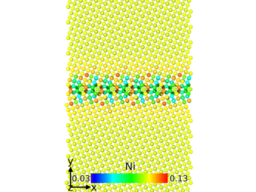}
\caption{}
\end{subfigure}
\begin{subfigure}[b]{0.24\textwidth}
\includegraphics[width=\textwidth]{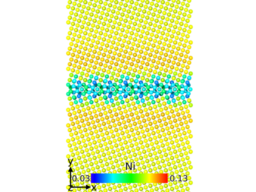}
\caption{}
\end{subfigure}
\begin{subfigure}[b]{0.24\textwidth}
\includegraphics[width=\textwidth]{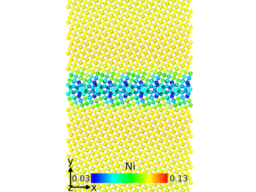}
\caption{}
\end{subfigure}
\caption {Evolution of Ni atomic molar fraction at the grain boundary in Cu-10at\%Ni alloy at $T =300$ K after (a) 0 second, (b) 0.2 second, (c) 2.20 seconds and (d) 8.20 seconds.} 
\label{Fig:CuNialloy-3}
\end{figure}

We study the effect of grain boundaries (GB) in Cu-10at\%Ni alloy at diverse temperatures. The model consists of two semi-infinite fcc grains with different misorientation angles that merges on a plane parallel to the $xz$ plane. The initial lattice parameter of the crystals is taken in function of the composition of the alloy and the temperature of the system, see Figure~\ref{Fig:CuNialloy-1}(a). The model of the GB consists of a total of 19,200 sites with periodic boundary conditions applied in all directions. Initially, we set every site with a composition of 90\% of Cu and 10\% of Ni. Then, employing a NPT ensemble, we apply zero pressure at constant temperature to the system, while we allow the system to evolve with time, governed by the kinetic law and the entropy of the system. The diffusive parameters of the kinetic law are taken as $\nu = 10^{10}$ s$^{-1}$ and $Q_m = 1.59$ eV. We notice that the activation energy is very high for this case, then a large diffusive time step in the simulations might be used at first glance because the diffusion process is very slow. The diffusive time step is also strongly dependent with the temperature of the system: the lower temperature the larger diffusive time step might be used in the simulations. Therefore, the computational cost required to reach a specified elapsed time is higher as temperature is increased. 


Ni atomic molar fractions near the GB are shown in Figure~\ref{Fig:CuNialloy-1}(b) for three different temperatures, i.e, $T=300$ K, $T=400$ K, $T=600$ K. The atomic configurations  after the annealing process at constant temperature are shown in Figure~\ref{Fig:CuNialloy-2}. The annealing time was of the order of a few seconds. From these results, we can observe that Ni was repelled from the GB and was spread homogeneously across the specimen. At low temperature, the diffusion of the atoms is slower than for the other two temperature. At $T=300$ K, high Ni concentration still remains localized near the grain boundaries after $t=8.2$ seconds. By way of contrast, at higher temperatures, Ni concentration varies near the GB, but far away the concentration has a constant value. 

Figure \ref{Fig:CuNialloy-3} shows the time sequence of Ni diffusion at $T=300$ K until $t =8.2$ seconds. The sequence shows that the transient is very long, and the sites near the GB experience first an increment in Ni concentration. As time went by, however, this increment of Ni tends to be homogeneously distributed in the bulk. The example shows a nice interplay between the atomic configurations, and the concentration of the alloy. In particular, this shows evidence that the concentration is highly dependent of the stress state. This is in line with the work of performed by Pellicer \textit{et al.} \cite{Pellicer:2011} for a Ni0.56Cu0.44 film annealed at $T=625$ K, where a segregation of Cu was observed near the GB.

\subsection{H diffusion in Pd nanowire}

We simulated the diffusion of H in a Pd nanowire at room temperature ($T=300$ K). The type of diffusion of H in Pd is mainly interstitial due to the atom size difference between H, or dopant atoms, and Pd, or host atoms; thus, substitutional diffusion has been excluded in this study. In substitutional-type diffusion, the dopant atoms substitute the host atoms of the lattice to diffuse, whereas dopant atoms use the interstitial sites of the lattice in interstitial-type diffusion. The fcc-Pd lattice has two types of interstitials: octahedral and tetrahedral. The octahedral sites in fcc-Pd posses a lower formation energy with 1.77 eV, versus 1.90 eV for the formation energy of the tetrahedral sites. Thus, H atoms prefer octahedral sites to be located rather than tetrahedral sites. We only considered the octahedral sites in this simulation, in line with previous studies \cite{sun2017acceleration,sun2019atomistic,sun2018long,sun2017atomistic}. The computational model consists of a nanowire of 9.4 nm of diameter and 58.8 nm of length, composed by two types of sites: the host sites, which correspond to the sites of a fcc-lattice occupied only by Pd atoms; and, the octahedral-type interstitials sites of the host fcc-lattice, occupied \emph{fully or partially} by H atoms. The lattice parameter of Pd was set to $a_0 = 0.3922$ nm. The longitudinal axis of the nanowire is aligned with the $\langle 001 \rangle$ crystallographic direction of fcc lattice. The simulation cell contained a total of 539,693 sites: 269,839 Pd sites and 269,854 interstitial sites. Free boundary conditions were set free in all directions. 

During the simulation, the host sites were set to be fully occupied by Pd atoms, the atomic molar fraction was always one, while the interstitial sites were allowed to be fully or partially occupied by H atoms. The atomic molar fraction of the interstitial atoms fluctuated between 0 (empty site) to 1 (fully occupied site) depending on the local environment. Initially, all interstitial sites were set to an atomic molar fraction of 0.05, except the sites of one side of the nanowire, which were set to 1. We assumed that these sites were always in contact with an infinite large H reservoir which introduced H atoms into the nanowire. We employed the EAM interatomic potential presented in \cite{zhou2008} to model the Pd-H system. The diffusive parameters of the kinetic law were taken as $\nu = 8.1 \times 10^{12}$ s$^{-1}$ and $Q_m = 0.294$ eV, that correspond to a diffusion coefficient of $D =7 \times 10^{-12}$ m$^{2} \cdot$s$^{-1}$ for H in Pd bulk in $\alpha$-phase at $T=250$ K \cite{Narayan2017}. Initially, we relaxed the system at a constant temperature of $T=300$ K by minimizing the energy. Then, we allowed the atomic molar fraction of the H sites to evolve with time and governed by the kinetic law and the free energy of the system. 

\begin{figure}
\centering
\begin{subfigure}[b]{0.32\textwidth}
\includegraphics[width=\textwidth]{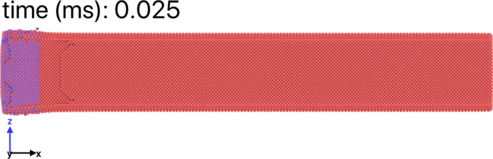}
\caption{}
\end{subfigure}
\begin{subfigure}[b]{0.32\textwidth}
\includegraphics[width=\textwidth]{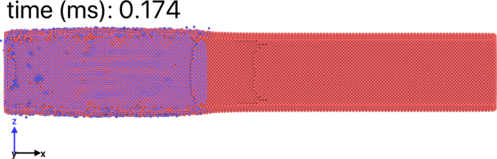}
\caption{}
\end{subfigure}
\begin{subfigure}[b]{0.32\textwidth}
\includegraphics[width=\textwidth]{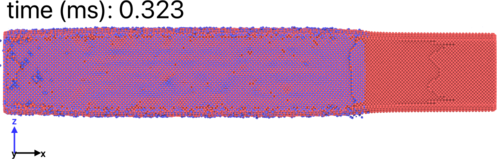}
\caption{}
\end{subfigure}
\caption {Simulation of the diffusion of H along the longitudinal axis of a Pd nanowire after: (a) 0.025 ms., (b) 0.174 ms. and (c) 0.323 ms. H atoms are shown in violet color
and Pd atoms in beige color. The size of the H atoms is scaled according to their atomic molar fraction.} 
\label{Fig:HPd-sim}
\end{figure}

\begin{figure}
\centering
\begin{subfigure}[b]{0.32\textwidth}
\includegraphics[width=\textwidth]{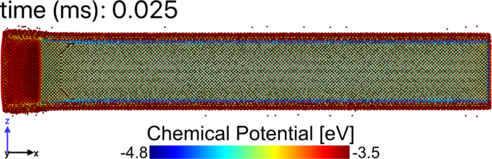}
\caption{}
\end{subfigure}
\begin{subfigure}[b]{0.32\textwidth}
\includegraphics[width=\textwidth]{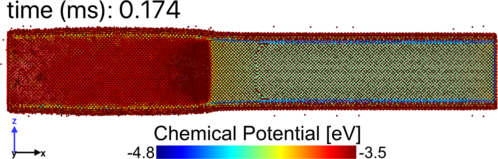}
\caption{}
\end{subfigure}
\begin{subfigure}[b]{0.32\textwidth}
\includegraphics[width=\textwidth]{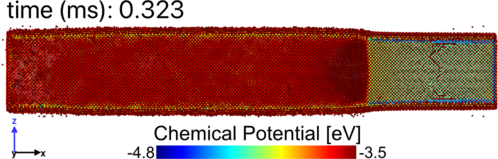}
\caption{}
\end{subfigure}
\caption {Evolution of the chemical potential during the diffusion of H in a Pd nanowire: (a) 0.025 ms., (b) 0.174 ms. and (c) 0.323 ms. }
\label{Fig:HPd-chemical}
\end{figure}

A time sequence of the diffusion of H in the Pd nanowire is shown in Figure~\ref{Fig:HPd-sim} where H atoms are decorated with violet color. As may be seen from the snapshot sequence, H gradually intercalated along the longitudinal axis of the Pd nanowire from domains with high concentration to domains with low H. From Figure~\ref{Fig:HPd-chemical}, we see that the domains where H has diffused, i.e., domains with high H concentration, the nanowire has larger values of the chemical potential in comparison with domains with low H concentration. The presence of this gradient of chemical potential generates a driving force that pushed atoms to diffuse to the areas of low H concentration in the nanowire. We also observed that the chemical potential between these two regions was always sharp. Similarly to previous works on H diffusion in Pd  \cite{sun2017acceleration,sun2019atomistic,sun2018long,sun2017atomistic}, our simulation also revealed that Pd undergoes a phase transition from $\alpha$-phase to $\beta$-phase. The $\beta$-phase preserved the fcc crystallographic structure; however, with a larger lattice parameter, which resulted in a diameter increase of the nanowire, as evidenced in Figures \ref{Fig:HPd-sim} and \ref{Fig:HPd-chemical}. Our results show that the lattice parameter of Pd $\beta-$phase increased up to approximately 0.4225 nm, i.e. an increment of 7.73$\%$. 

\begin{figure}
\centering
\begin{subfigure}[b]{0.45\textwidth}
\includegraphics[width=\textwidth]{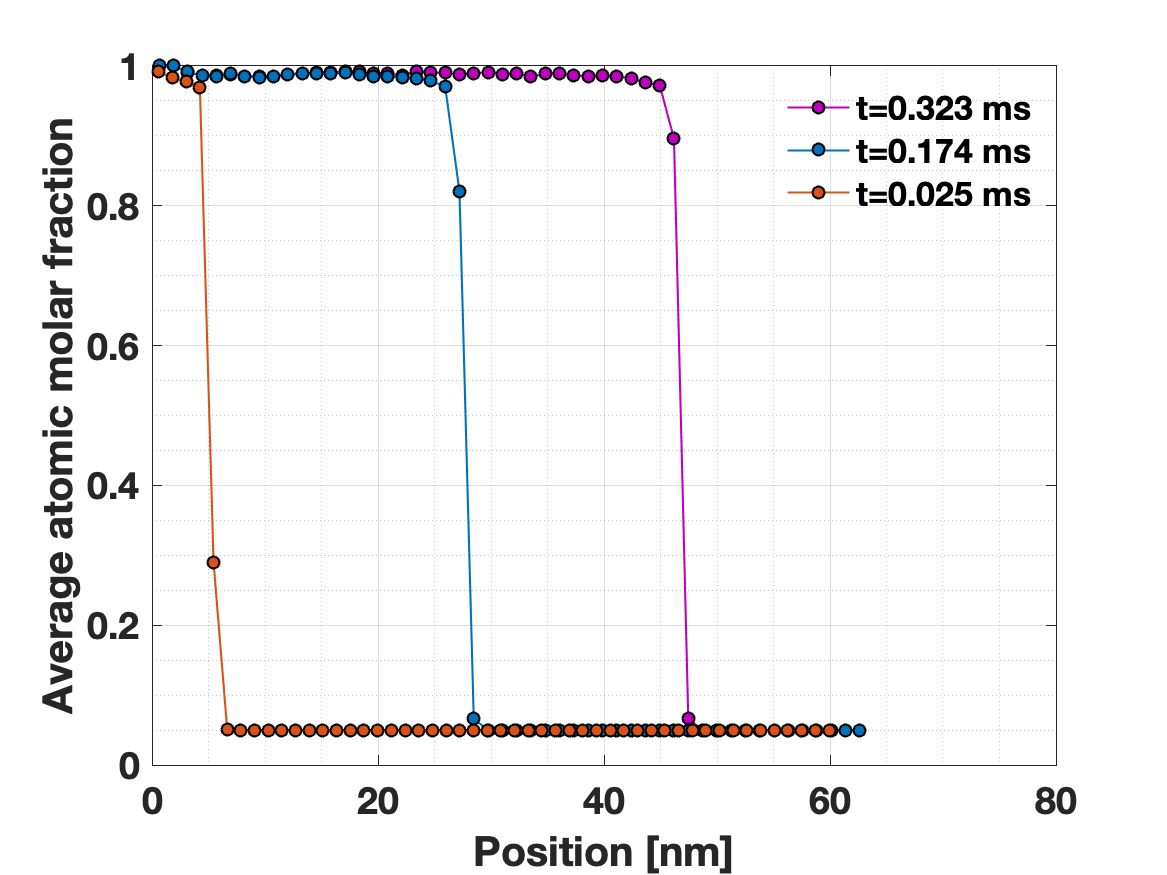}
\caption{}
\end{subfigure}
\begin{subfigure}[b]{0.45\textwidth}
\includegraphics[width=\textwidth]{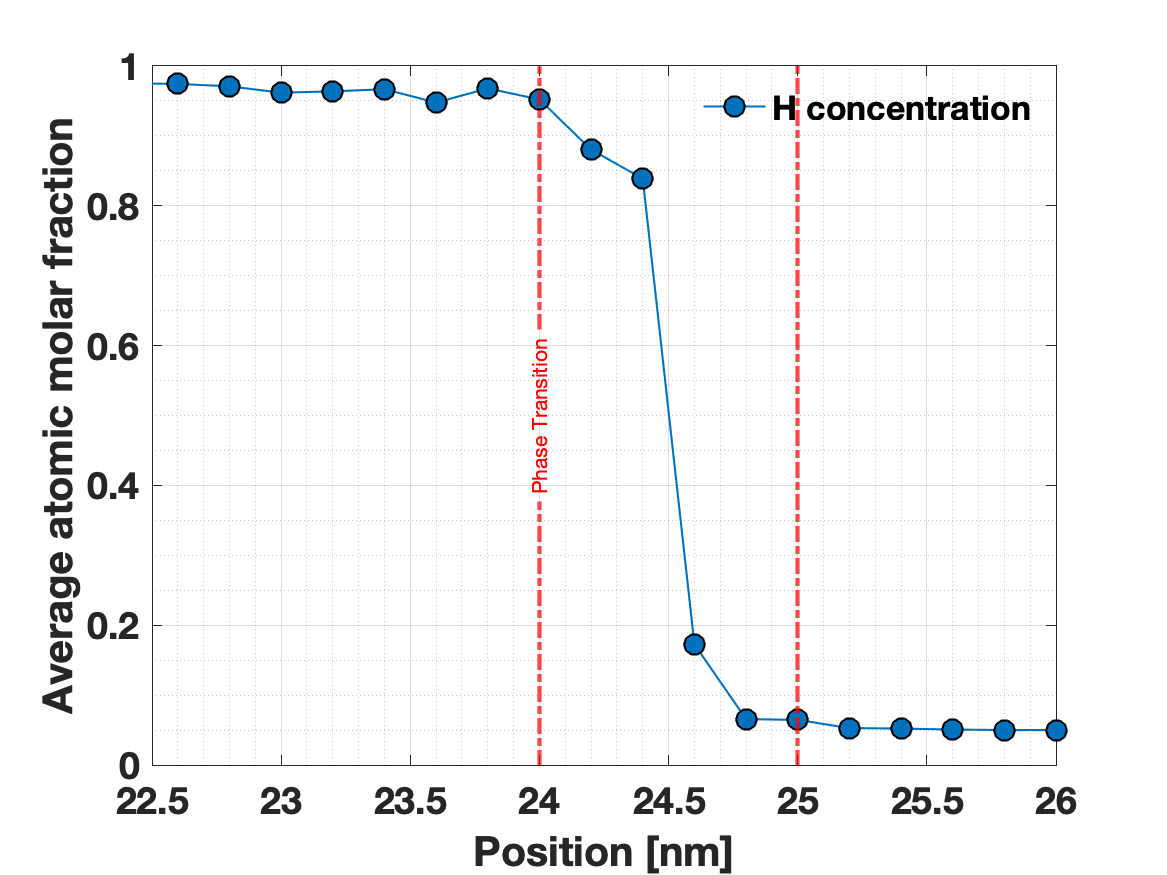}
\caption{}
\end{subfigure}
\caption {Spatial and time evolution of average  atomic molar fraction for H in a Pd nanowire for (a) multiple times, and (b) phase boundary after $t = 0.174$ ms. } 
\label{Fig:HPd-phase_boundary}
\end{figure}

In systems that undergo phase transformations, the mass transport often occurs by the motion of a sharp phase boundary. Our simulation predicted such phase boundary, and revealed that the thickness of the phase boundary is of the order of $\sim1.0$ nm, see Figure~\ref{Fig:HPd-phase_boundary}, in good agreement with previous studies \cite{sun2017acceleration,sun2019atomistic,sun2018long,sun2017atomistic} of Pd nanoparticles. Figure~\ref{Fig:HPd-phase_boundary}(a) shows the motion of the phase boundary at different times in the simulation. We see that for the three selected times, a sharp interface is always present. Figure~\ref{Fig:HPd-phase_boundary}(b) shows a close-up of the concentration at $t=0.173$ ms, where the interface is indicated with two red dashed lines. 

\section{Parallel performance}\label{Sec:Performance}

In high performance computing, a good scalability and efficiency of the code are crucial to perform efficiently large-scale simulations. Using the vacancy diffusion in fcc-Cu from Section~\ref{Sec:Applications:Vacancy_diffusion} as the test case, we have measured the strong and weak scaling and the corresponding efficiencies for a wide range of number of processors, up to 512 processors. The \emph{strong scaling} shows how the computational time varies with the number of processors for a total fixed load or number of atoms, while the \emph{weak scaling} represents how the computational time changes for a fixed load size per processor (number of atoms per processor is fixed). The simulations have been run on MPI-cluster with 16 cores/socket, 2 sockets/node, each core is an Intel$^{\textregistered}$ Broadwell$^{\textregistered}$ CPUs, model E5-2683 v4, which have a clock speed of 2.1 GHz. 

\begin{figure}
\centering
\begin{subfigure}[b]{0.45\textwidth}
\includegraphics[width=\textwidth]{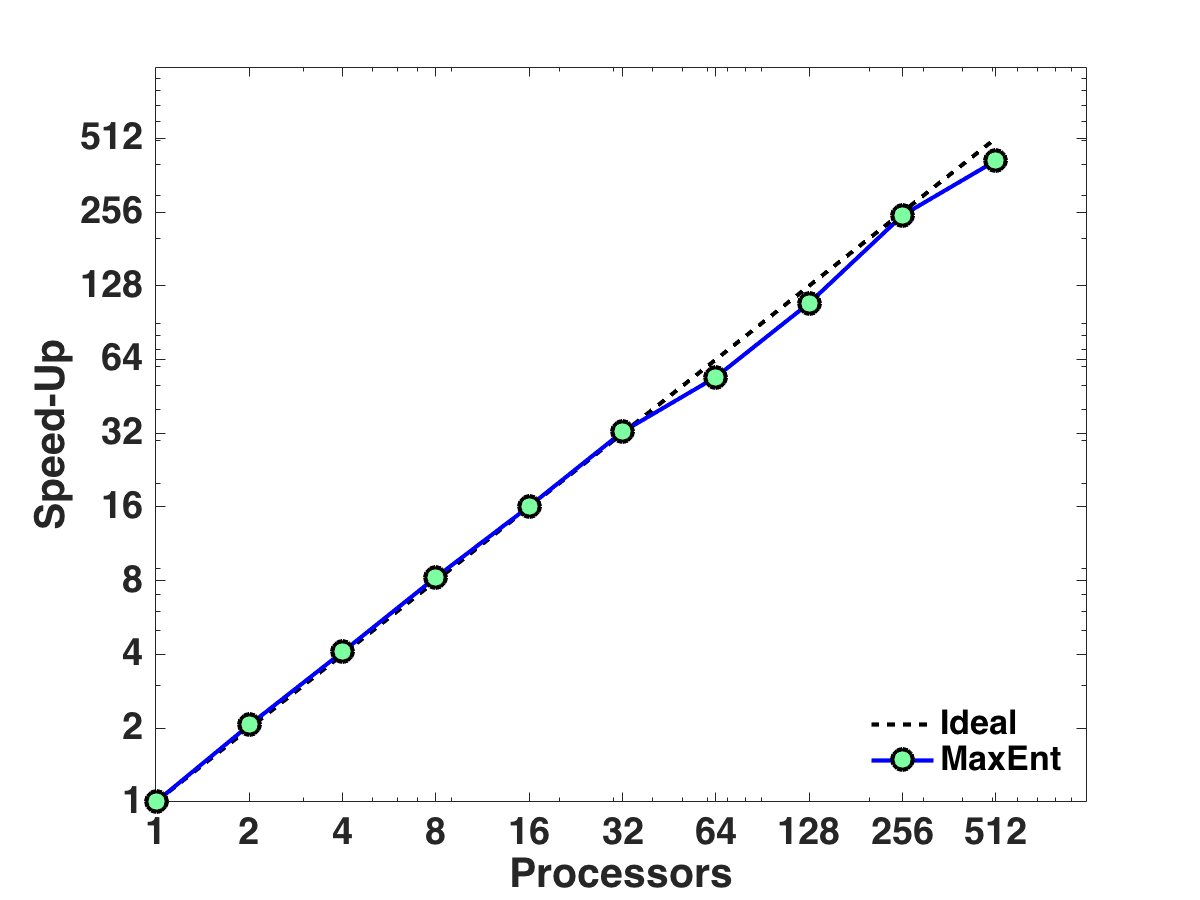}
\caption{}
\end{subfigure}
\begin{subfigure}[b]{0.45\textwidth}
\includegraphics[width=\textwidth]{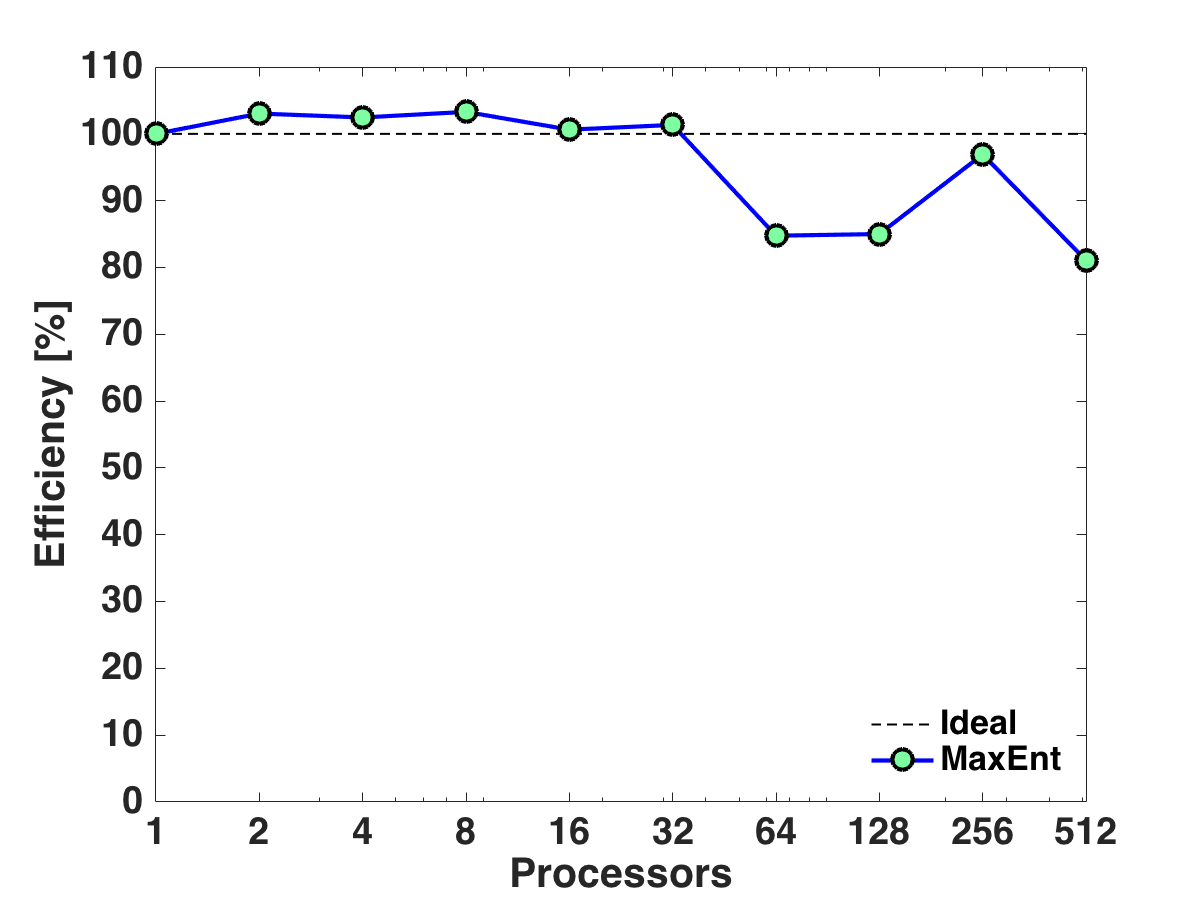}
\caption{}
\end{subfigure}
\caption{Strong scaling speed-up and efficiency of the MXE package in LAMMPS for a computational cell containing a total of 864,000 atoms.}
\label{Speed-Up}
\end{figure}

\begin{figure}
\centering
\begin{subfigure}[b]{0.45\textwidth}
\includegraphics[width=\textwidth]{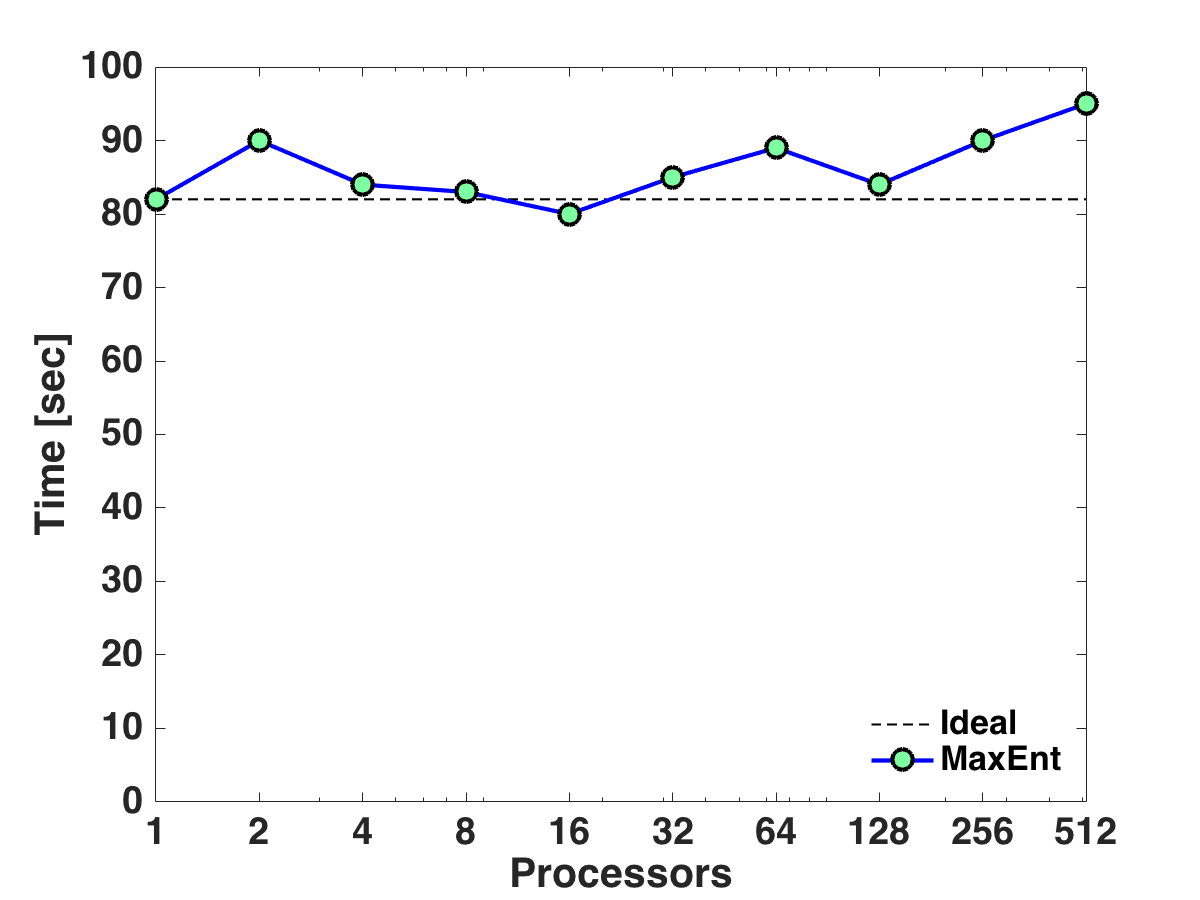}
\caption{}
\end{subfigure}
\begin{subfigure}[b]{0.45\textwidth}
\includegraphics[width=\textwidth]{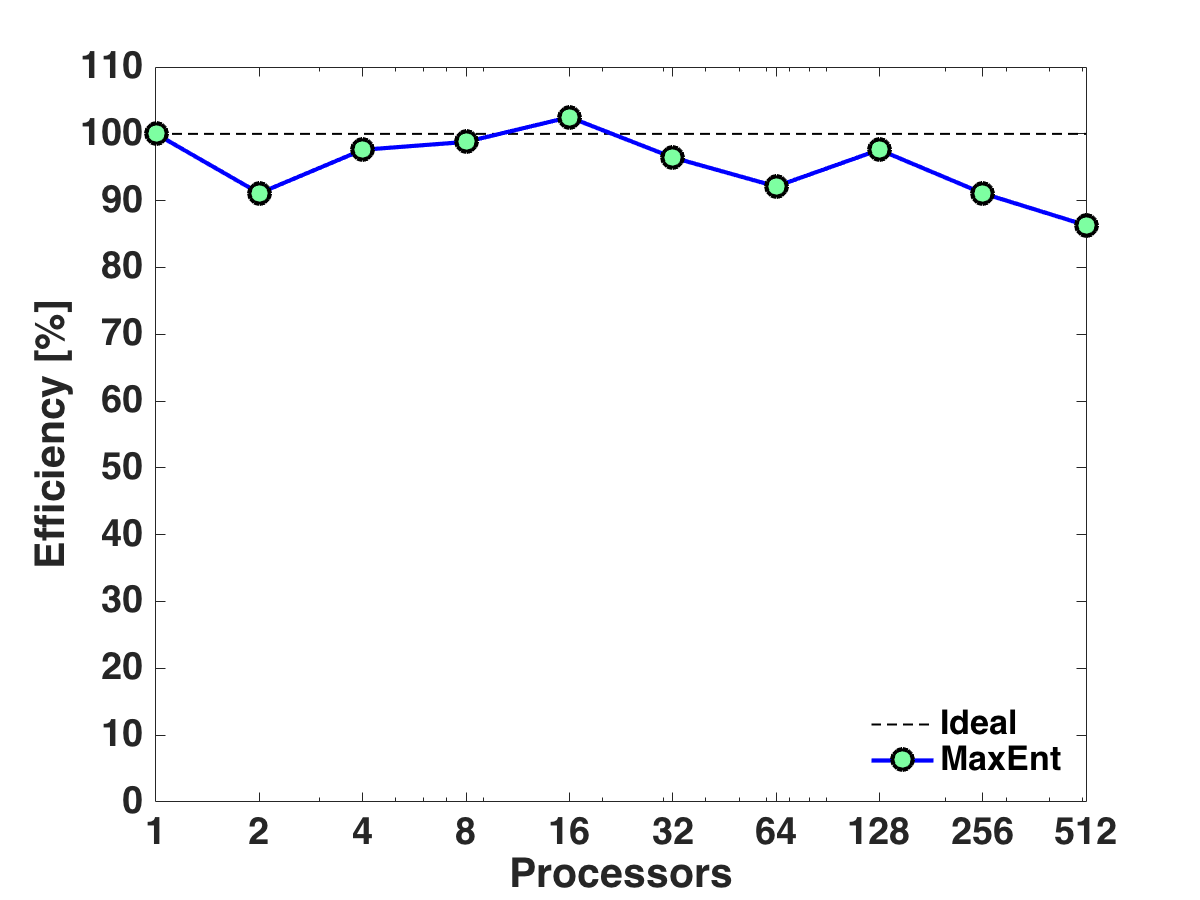}
\caption{}
\end{subfigure}
\caption{Weak scaling performance and efficiency of the MXE package in LAMMPS for a computational cell containing 32,000 atoms per processor.}
\label{WeakScaling}
\end{figure}

Regarding the strong scaling, we measured the relative speed-up as $\frac{t_{N_{proc}=1}}{t_{N_{proc}}}$, where $t_{N_{proc}}$ is the time to complete the simulation with $N_{proc}$. Figure~\ref{Speed-Up} shows the speed-up, together with the corresponding efficiency, for a computational cell size of 864,000 atoms. The obtained speed-up is very close to the ideal and the efficiency is around 81\% for $N_{proc} = 512$. Next, Figure \ref{WeakScaling} shows the weak scaling for the simulation case. The plot indicates the computational time for a giving number of processors required to fully complete one hundred evaluations of forces per atom with a constant load of 32,000 atoms per processor. As shown, the required computational time and the efficiency were kept constant with the number of processors, giving an efficiency of around 86\% for 512 processors (the simulation cell contained a total of 16,384,000 atoms). The outstanding performance of the MXE package is mainly due to the fact that it is implemented as library on top of LAMMPS, which is highly efficient on high-performance clusters. 

While the implementation of mxe framework involves some extra overheads with respect to the original LAMMPS code, Figure \ref{Speed-Up} indicates that the performance of the new implementation is not strongly deteriorated by these overheads. This extra computational time is the minimum price that we need to be paid to achieve the tremendous acceleration in the simulation time. We can distinguish two types of overheads. The first one includes the communication of new quantities such as atomic temperature, frequency, atomic molar fraction and forces between processor during the simulations. The second one is the evaluation of the phase-average of the energies and forces, which is more computational expensive than the counterpart in MD. This, in turn, increases the ratio between computation and communication among processors, leading to a very minor penalty in the performance of the code with respect to original LAMMPS code, see Figure \ref{Speed-Up}. In theory, the estimated overhead using third-degree Gauss-Hermite quadrature rules is around 12 times higher than a regular force evaluation. However, this overhead is normally exceeded in practice due to some extra calculations (e.g. the computation of the  forces with respect to frequency, among others) and extra communications between processors during the computation of the phase averages.

\section{Conclusions}\label{Sec:Conclusions}

We have implemented and validated a package to simulate long term diffusive phenomena in atomistic systems. The user package was implemented as an add-on to the popular and open source LAMMPS code. The implementation is flexible, efficient, user-friendly, and compatible with other LAMMPS' features. The package extends the realm of MD simulations to diffusive problems that cannot be achieved with the \emph{state-of-the-art} \emph{traditional} molecular dynamics techniques. 

We have shown the simulation of equilibrium and non-equilibrium problems in a variety of materials, illustrating the flexibility of the methodology and the implementation to model complex systems where mass transport is of interest. 
In many examples, comparison with either analytical solutions, MD, MC, and/or experiments has been carried out, with good agreement in all cases. As such, the current framework could serve as the basis for the next generation of mixed empirical-theoretical models, where a large part of the information is taken from \textit{ab-initio} techniques --- such as forces and atomic interactions --- while the transport laws and coefficients can be calibrated from experimental measurements. This trade-off between theory and experiments is often desired in applications where a large number of parameters are unknown or uncontrolled.

Possible extensions of the method rely on the proposition of more-sophisticated kinematic transport laws, and better local trial Hamiltonians that can provide controlled approximation schemes. The diffusive framework can also be applied to other methodologies that include particle-based methods at the continuum scale such as Peridynamics and the  Smooth Particle Hydrodynamics methods. It is also noteworthy that the framework is amenable to spatial coarse-graining by recourse of the quasi-continuum method; this final task is an avenue that the authors are actively pursuing.

\section{Acknowledgments}

We gratefully acknowledge the support from the Natural Sciences and Engineering Research Council of Canada (NSERC) through the Discovery Grant under Award Application Number RGPIN-2016-06114 and the support of Compute Canada through the Westgrid consortium for giving access to the supercomputer grid. 

\begin{appendix}

\section{Evaluation of the phase average} \label{Appendix:Sec:phase_average}

Within this framework, the evaluation of energies and forces needs to be computed by the phase average against the mean field probability density function $p_0$, see Equation~\ref{Eq:MeanFieldRho}. In the MXE package, we have used the third-degree Gauss-Hermite quadrature rules \cite{Stroud:1971} to evaluate efficiently these phase averages. 

We begin by considering an arbitrary $N-$body interaction potential of dimension $N_q = 3N$, $\phi(\textbf{q}_1, \textbf{q}_2, \ldots, \textbf{q}_N)$. The expected value, or phase average, is computed as
\begin{align} \label{Appendix:Eq:phase_average}
\langle \phi(\textbf{q}_1, \ldots, \textbf{q}_N) \rangle_0 & =  \int_{\Gamma} \phi(\textbf{q}_1, \ldots, \textbf{q}_N) \prod_{i=1}^N p_{0i} d\textbf{q}_i \\ \nonumber
& = \int_{\Gamma} \phi(\textbf{q}_1, \ldots, \textbf{q}_N) \prod_{i=1}^N \dfrac{1}{Z_0} \exp(\frac{-\beta_i m_i \omega_i^2}{2 } \vert \textbf{q}_i - \overline{\textbf{q}}_i \vert^2) d\textbf{q}_i \\ \nonumber
& = \left( \frac{1}{\sqrt{\pi}} \right)^{3N} \int_{\infty}^{\infty} \ldots  \int_{\infty}^{\infty} \phi ( \boldsymbol \chi_1, \ldots, \boldsymbol \chi_N) \exp(-\vert {\boldsymbol \chi}_1 \vert^2 \ldots -\vert {\boldsymbol \chi}_N \vert^2) d{\boldsymbol \chi}_1 \ldots d{\boldsymbol \chi}_N, 
\end{align}
where $\langle \rangle_0$ refers to the phase average against $\rho_0$ and
\begin{equation}
{\boldsymbol \chi}_i = \sqrt{\dfrac{\beta_i m_i}{2}}\omega_i (\textbf{q}_i  - \overline{\textbf{q}}_i).
\end{equation}
A M-point numerical quadrature approximates the phase average of $\phi(\textbf{q}_1, \textbf{q}_2, \ldots, \textbf{q}_N)$ as
\begin{equation} \label{Appendix:Eq:QuadRule}
\langle \phi (\textbf{q}_1, \ldots, \textbf{q}_N) \rangle \approx \left( \frac{1}{\sqrt{\pi}} \right)^{3N} \sum_{k=1}^M  \phi ({\boldsymbol \xi}_k) W_k,
\end{equation}
where $M = 2 \times N_q$ is the number of quadrature points, $W_k$ is the corresponding weight, and ${\boldsymbol \xi}_k$ is the $k^{th}$ component of the $N_{q}$-dimensional subarray of the multidimensional array
\begin{equation}
\boldsymbol \xi = 
\bigg\{ 
\big\{ \begin{pmatrix}r\\0\\0\end{pmatrix},..., \begin{pmatrix}0\\0\\0\end{pmatrix} \big\},
\big\{ \begin{pmatrix}-r\\0\\0\end{pmatrix},..., \begin{pmatrix}0\\0\\0\end{pmatrix} \big\},
\big\{ \begin{pmatrix}0\\r\\0\end{pmatrix},..., \begin{pmatrix}0\\0\\0\end{pmatrix} \big\},...,
\big\{ \begin{pmatrix}0\\0\\0\end{pmatrix},..., \begin{pmatrix}0\\0\\-r\end{pmatrix} \big\}
\bigg\},
\end{equation}
whose columns give the position of the nodes at the $k^{th}$ quadrature node. The values of the weights are given by
\begin{equation}
W_k = \frac{\pi^{N_q/2}}{2N_q},\  k=1,\ldots,M
\end{equation}
with
\begin{equation}
r^2 = \frac{N_q}{2}.
\end{equation}

The nodes and weights are obtained by requiring that the formula should integrate exactly all polynomials of degree $\le 3$. Since most of the potentials are given by pice-wise continuous cubic spline, the quadrature rule of third order is very convenient, because it gives a good balance between efficiency and accuracy.

\section{Thermalization of interatomic potentials}\label{Appendix:Sec:Potentials}

We present the calculation of the phase average of several potentials. We first start with a very simple potential such as the Lennard-Jones potential in \ref{Appendix:Sec:Potentials:LJ}. Similar potentials, such as Stillinger and Webber potential, are thermalized in a very similar way. Then, we increase the complexity of the phase average of potentials with the thermalization of the EAM potential in \ref{Appendix:Sec:Potentials:EAM}.

\subsection{Lennard-Jones potential}\label{Appendix:Sec:Potentials:LJ}

The 12/6 Lennard-Jones (LJ) potential is given by 
\begin{equation}\label{Appendix:Eq:LJ}
V=\sum_{i=1}^N \sum_{\substack{j=1 \\ j \leq i}}^N \phi_{ij} = \sum_{i=1}^N \sum_{\substack{j=1 \\ j \leq i}}^N 4 \epsilon \bigg[ \bigg( \frac{\sigma}{r_{ij}} \bigg)^{12} - \bigg(\frac{\sigma}{r_{ij}} \bigg)^6 \bigg]
\end{equation}
where $r_{ij}=|{\bf r}_i-{\bf r}_j|$, and $ \epsilon $ and $\sigma$ are parameters of the potential.

The phase average of the LJ potential is then given by 
\begin{equation}\label{Appendix:Eq:LJ_mxe}
\langle V \rangle_0 =\sum_{i=1}^N \sum_{\substack{j=1 \\ j \leq i}}^N \langle \phi_{ij} \rangle_0 = \sum_{i=1}^N \sum_{\substack{j=1 \\ j \leq i}}^N 4 \epsilon \langle \bigg( \frac{\sigma}{r_{ij}} \bigg)^{12} - \bigg(\frac{\sigma}{r_{ij}} \bigg)^6 \rangle_0.
\end{equation}
The right hand side only involves pair-wise interactions and its thermalization using third-order Gauss-Hermite quadrature is straightforward. The evaluation of the phase average using third degree Gauss-Hermite quadrature rule is presented in details in \ref{Appendix:Sec:phase_average}.

\subsection{Embedded atom model potential}\label{Appendix:Sec:Potentials:EAM}

The embedded atom model (EAM) potential energy is written as \citep{Daw:1984}
\begin{equation} \label{Appendix:Eq:EAM}
V =  \sum_{i=1}^N \sum_{\substack{j=1 \\ j \leq i}}^N \phi(r_{ij}) + \sum_{i=1}^N F(\rho_i) 
\end{equation}
where
\begin{equation} \label{Appendix:Eq:EAM_density}
\rho_i = \sum_{j=1}^N f(r_{ij})
\end{equation}
is the electron density at atom $i$ due to neighbor atoms $j$ and $r_{ij} = |{\bf r}_i - {\bf r}_j |$. 

The phase average of Equation \ref{Appendix:Eq:EAM} is given by
\begin{equation}\label{Appendix:Eq:EAM_mxe}
\langle V \rangle_0 =  \sum_{i=1}^N \sum_{\substack{j=1 \\ j \leq i}}^N \langle \phi(r_{ij}) \rangle_0 + \sum_{i=1}^N \langle F(\rho_i) \rangle_0,
\end{equation}
where the first term in \ref{Appendix:Eq:EAM_mxe} only involves pair-wise interactions and its corresponding phase average calculation is straightforward, see \ref{Appendix:Sec:phase_average} and \ref{Appendix:Sec:Potentials:LJ}. On the contrary, the second term, the embedding function $F(\rho_i)$, includes many body interactions through the term $\rho_i$ and, therefore, the phase average is more involved. The number of quadrature nodes -and potential evaluation-, see \ref{Appendix:Sec:phase_average}, scales proportionally with the number of neighbor atoms inside the potential cut-off radius. Thus, the evaluation of the phase average is computationally intense even in the case of EAM potentials with short cut-off radius. Furthermore, the more neighbor atoms the more numerical error we might incur in our calculations, since we are approximating a function of higher dimension with a third order quadrature rule. In order to make calculations more efficient and reduce the overhead with respect to regular MD, we simplify the calculation of the phase average expanding the embedding function as a Taylor series near an equilibrium electronic density

\begin{equation} \label{Appendix:Eq:F_taylor}
F_i(\rho_i) \approx F_i(\rho_i|_0) + F_i'(\rho_i|_0)(\rho_i-\rho_i|_0)
\end{equation}
where $\rho_i|_0$ refers to the electron density evaluated at the mean positions $\bar{\textbf{q}}_i$. Then, the phase average of Equation \ref{Appendix:Eq:F_taylor} is giving by
\begin{equation} \label{Appendix:Eq:F_taylor_mxe}
\langle F_i(\rho_i) \rangle_0 \approx F_i(\rho_i|_0) +  F_i'(\rho_i|_0)( \langle \rho_i \rangle_0 -\rho_i|_0)
\end{equation}
This approach has been successfully used in the past \cite{LeSar:1991, Venturini:2014, Mendez:2018}.

\subsubsection{EAM potential for alloys within this framework}

We modify the EAM potential energy, Equation \ref{Appendix:Eq:EAM}, including the occupancy function as
\begin{equation} \label{EAM-Substitutional}
V_i =  \frac{1}{2} \sum_{j=1}^N \sum_{k=1}^M \sum_{l=1}^M n_{ik} n_{jl}  \phi_{kl}(r_{ij})  + \sum_{k=1}^M n_{ik}  F_k(\rho_i) 
\end{equation}
where the electron density at site $i$ is
\begin{equation} \label{EAM-density-substitutional}
\rho_i = \sum_{j=1}^N \sum_{l=1}^M n_{jl}  f(r_{ij}),
\end{equation}
and $M$ is the total number of species in the system.
Following \cite{Venturini:2014, Mendez:2018}, we approximate the phase average of \ref{EAM-Substitutional} and \ref{EAM-density-substitutional} as
\begin{equation} \label{EAM-Substitutional-ave}
\langle V_i \rangle_0 \approx \frac{1}{2} \sum_{j=1}^N \sum_{k=1}^M \sum_{l=1}^M x_{ik} x_{jl} \langle \phi_{kl}(r_{ij}) \rangle_0 + \sum_{k=1}^M x_{ik} \langle F_k(\rho_i) \rangle_0
\end{equation}
and
\begin{equation} \label{EAM-density-substitutional-ave}
\langle \rho_i \rangle_0 \approx \sum_{j=1}^N \sum_{l=1}^M x_{jl}  \langle f(r_{ij}) \rangle_0,
\end{equation}
respectively, where $x_{ik}$ is the atomic molar fraction of specie $k$ at site $i$. Similarly, the phase average of the embedding function is treated in a similar way as described above for pair potentials.

\end{appendix}

\section*{References}
\bibliography{myreferences}

\begin{thebibliography}{56}
\expandafter\ifx\csname natexlab\endcsname\relax\def\natexlab#1{#1}\fi
\providecommand{\bibinfo}[2]{#2}
\ifx\xfnm\relax \def\xfnm[#1]{\unskip,\space#1}\fi
\bibitem[{Bardeen(1949)}]{Bardeen:1949}
\bibinfo{author}{J.~Bardeen},
\newblock \bibinfo{title}{Diffusion in binary alloys},
\newblock \bibinfo{journal}{Phys. Rev.} \bibinfo{volume}{76}
  (\bibinfo{year}{1949}) \bibinfo{pages}{1403--1405}.
\bibitem[{Elcock and McCombie(1958)}]{Elcock:1958}
\bibinfo{author}{E.~W. Elcock}, \bibinfo{author}{C.~W. McCombie},
\newblock \bibinfo{title}{Vacancy diffusion in binary ordered alloys},
\newblock \bibinfo{journal}{Phys. Rev.} \bibinfo{volume}{109}
  (\bibinfo{year}{1958}) \bibinfo{pages}{605--606}.
\bibitem[{Manning(1967)}]{Manning:1967}
\bibinfo{author}{J.~Manning},
\newblock \bibinfo{title}{Diffusion and the kirkendall shift in binary alloys},
\newblock \bibinfo{journal}{Acta Metallurgica} \bibinfo{volume}{15}
  (\bibinfo{year}{1967}) \bibinfo{pages}{817 -- 826}.
\bibitem[{Matan et~al.(1999)Matan, Cox, Carter, Rist, Rae, and
  Reed}]{Creep:1999}
\bibinfo{author}{N.~Matan}, \bibinfo{author}{D.~Cox},
  \bibinfo{author}{P.~Carter}, \bibinfo{author}{M.~Rist},
  \bibinfo{author}{C.~Rae}, \bibinfo{author}{R.~Reed},
\newblock \bibinfo{title}{Creep of cmsx-4 superalloy single crystals: effects
  of misorientation and temperature},
\newblock \bibinfo{journal}{Acta Materialia} \bibinfo{volume}{47}
  (\bibinfo{year}{1999}) \bibinfo{pages}{1549 -- 1563}.
\bibitem[{Reed(2006)}]{Reed:2008}
\bibinfo{author}{R.~C. Reed}, \bibinfo{title}{The Superalloys: Fundamentals and
  Applications}, \bibinfo{number}{v. 1}, \bibinfo{publisher}{Cambridge
  University Press}, \bibinfo{year}{2006}.
\bibitem[{Haghighat et~al.(2013)Haghighat, Eggeler, and Raabe}]{Climb:2013}
\bibinfo{author}{S.~H. Haghighat}, \bibinfo{author}{G.~Eggeler},
  \bibinfo{author}{D.~Raabe},
\newblock \bibinfo{title}{Effect of climb on dislocation mechanisms and creep
  rates in Î³â²-strengthened ni base superalloy single crystals: A
  discrete dislocation dynamics study},
\newblock \bibinfo{journal}{Acta Materialia} \bibinfo{volume}{61}
  (\bibinfo{year}{2013}) \bibinfo{pages}{3709 -- 3723}.
\bibitem[{Chen(2005)}]{NanoscaleEnergy:2005}
\bibinfo{author}{G.~Chen}, \bibinfo{title}{Nanoscale energy transport and
  conversion : a parallel treatment of electrons, molecules, phonons, and
  photons}, \bibinfo{publisher}{Oxford ; New York : Oxford University Press,
  2005.}, \bibinfo{year}{2005}.
\bibitem[{Pop(2010)}]{Pop:2010}
\bibinfo{author}{E.~Pop},
\newblock \bibinfo{title}{Energy dissipation and transport in nanoscale
  devices},
\newblock \bibinfo{journal}{Nano Research} \bibinfo{volume}{3}
  (\bibinfo{year}{2010}) \bibinfo{pages}{147--169}.
\bibitem[{Kelly and Miller(2007)}]{Kelly:2007}
\bibinfo{author}{T.~F. Kelly}, \bibinfo{author}{M.~K. Miller},
\newblock \bibinfo{title}{Atom probe tomography},
\newblock \bibinfo{journal}{Review of Scientific Instruments}
  \bibinfo{volume}{78} (\bibinfo{year}{2007}) \bibinfo{pages}{031101}.
\bibitem[{Miller et~al.(2007)Miller, Russell, Thompson, Alvis, and
  Larson}]{Miller:2007}
\bibinfo{author}{M.~K. Miller}, \bibinfo{author}{K.~F. Russell},
  \bibinfo{author}{K.~Thompson}, \bibinfo{author}{R.~Alvis},
  \bibinfo{author}{D.~J. Larson},
\newblock \bibinfo{title}{Review of atom probe fib-based specimen preparation
  methods},
\newblock \bibinfo{journal}{Microscopy and Microanalysis} \bibinfo{volume}{13}
  (\bibinfo{year}{2007}) \bibinfo{pages}{428–436}.
\bibitem[{Kelly and Larson(2012)}]{Kelly:2012}
\bibinfo{author}{T.~F. Kelly}, \bibinfo{author}{D.~J. Larson},
\newblock \bibinfo{title}{Atom probe tomography 2012},
\newblock \bibinfo{journal}{Annual Review of Materials Research}
  \bibinfo{volume}{42} (\bibinfo{year}{2012}) \bibinfo{pages}{1--31}.
\bibitem[{Voter(1997)}]{HyperMD}
\bibinfo{author}{A.~F. Voter},
\newblock \bibinfo{title}{Hyperdynamics: Accelerated molecular dynamics of
  infrequent events},
\newblock \bibinfo{journal}{Phys. Rev. Lett.} \bibinfo{volume}{78}
  (\bibinfo{year}{1997}) \bibinfo{pages}{3908--3911}.
\bibitem[{Sorensen and Voter(2000)}]{TAD}
\bibinfo{author}{M.~R. Sorensen}, \bibinfo{author}{A.~F. Voter},
\newblock \bibinfo{title}{Temperature-accelerated dynamics for simulation of
  infrequent events},
\newblock \bibinfo{journal}{The Journal of Chemical Physics}
  \bibinfo{volume}{112} (\bibinfo{year}{2000}) \bibinfo{pages}{9599--9606}.
\bibitem[{Perez et~al.(2009)Perez, Uberuaga, Shim, Amar, and
  Voter}]{Perez:2009}
\bibinfo{author}{D.~Perez}, \bibinfo{author}{B.~P. Uberuaga},
  \bibinfo{author}{Y.~Shim}, \bibinfo{author}{J.~G. Amar},
  \bibinfo{author}{A.~F. Voter},
\newblock \bibinfo{title}{Chapter 4 accelerated molecular dynamics methods:
  Introduction and recent developments},
\newblock volume~\bibinfo{volume}{5} of \textit{\bibinfo{series}{Annual Reports
  in Computational Chemistry}}, \bibinfo{publisher}{Elsevier},
  \bibinfo{year}{2009}, pp. \bibinfo{pages}{79 -- 98}.
\bibitem[{Sadigh et~al.(2012)Sadigh, Erhart, Stukowski, Caro, Martinez, and
  Zepeda-Ruiz}]{MC:2012}
\bibinfo{author}{B.~Sadigh}, \bibinfo{author}{P.~Erhart},
  \bibinfo{author}{A.~Stukowski}, \bibinfo{author}{A.~Caro},
  \bibinfo{author}{E.~Martinez}, \bibinfo{author}{L.~Zepeda-Ruiz},
\newblock \bibinfo{title}{Scalable parallel monte carlo algorithm for atomistic
  simulations of precipitation in alloys},
\newblock \bibinfo{journal}{Phys. Rev. B} \bibinfo{volume}{85}
  (\bibinfo{year}{2012}) \bibinfo{pages}{184203}.
\bibitem[{Sadigh and Erhart(2012)}]{MC:2012b}
\bibinfo{author}{B.~Sadigh}, \bibinfo{author}{P.~Erhart},
\newblock \bibinfo{title}{Calculation of excess free energies of precipitates
  via direct thermodynamic integration across phase boundaries},
\newblock \bibinfo{journal}{Phys. Rev. B} \bibinfo{volume}{86}
  (\bibinfo{year}{2012}) \bibinfo{pages}{134204}.
\bibitem[{Mianroodi et~al.(2019)Mianroodi, Shanthraj, Kontis, Cormier, Gault,
  Svendsen, and Raabe}]{Mianroodi:2019}
\bibinfo{author}{J.~R. Mianroodi}, \bibinfo{author}{P.~Shanthraj},
  \bibinfo{author}{P.~Kontis}, \bibinfo{author}{J.~Cormier},
  \bibinfo{author}{B.~Gault}, \bibinfo{author}{B.~Svendsen},
  \bibinfo{author}{D.~Raabe},
\newblock \bibinfo{title}{Atomistic phase field chemomechanical modeling of
  dislocation-solute-precipitate interaction in ni–al–co},
\newblock \bibinfo{journal}{Acta Materialia} \bibinfo{volume}{175}
  (\bibinfo{year}{2019}) \bibinfo{pages}{250 -- 261}.
\bibitem[{Li et~al.(2011)Li, Sarkar, Cox, Lenosky, Bitzek, and
  Wang}]{DMD-Li:2011}
\bibinfo{author}{J.~Li}, \bibinfo{author}{S.~Sarkar}, \bibinfo{author}{W.~T.
  Cox}, \bibinfo{author}{T.~J. Lenosky}, \bibinfo{author}{E.~Bitzek},
  \bibinfo{author}{Y.~Wang},
\newblock \bibinfo{title}{Diffusive molecular dynamics and its application to
  nanoindentation and sintering},
\newblock \bibinfo{journal}{Phys. Rev. B} \bibinfo{volume}{84}
  (\bibinfo{year}{2011}) \bibinfo{pages}{054103}.
\bibitem[{Venturini et~al.(2014)Venturini, Wang, Romero, Ariza, and
  Ortiz}]{Venturini:2014}
\bibinfo{author}{G.~Venturini}, \bibinfo{author}{K.~Wang},
  \bibinfo{author}{I.~Romero}, \bibinfo{author}{M.~Ariza},
  \bibinfo{author}{M.~Ortiz},
\newblock \bibinfo{title}{Atomistic long-term simulation of heat and mass
  transport},
\newblock \bibinfo{journal}{Journal of the Mechanics and Physics of Solids}
  \bibinfo{volume}{73} (\bibinfo{year}{2014}) \bibinfo{pages}{242 -- 268}.
\bibitem[{Gonzalez-Ferreiro et~al.(2016)Gonzalez-Ferreiro, Romero, and
  Ortiz}]{Romero:2016}
\bibinfo{author}{B.~Gonzalez-Ferreiro}, \bibinfo{author}{I.~Romero},
  \bibinfo{author}{M.~Ortiz},
\newblock \bibinfo{title}{A numerical method for the time coarsening of
  transport processes at the atomistic scale},
\newblock \bibinfo{journal}{Modelling and Simulation in Materials Science and
  Engineering} \bibinfo{volume}{24} (\bibinfo{year}{2016})
  \bibinfo{pages}{045011}.
\bibitem[{Sun et~al.(2017)Sun, Ariza, Ortiz, and Wang}]{Sun:2017}
\bibinfo{author}{X.~Sun}, \bibinfo{author}{M.~Ariza},
  \bibinfo{author}{M.~Ortiz}, \bibinfo{author}{K.~Wang},
\newblock \bibinfo{title}{Acceleration of diffusive molecular dynamics
  simulations through mean field approximation and subcycling time
  integration},
\newblock \bibinfo{journal}{Journal of Computational Physics}
  \bibinfo{volume}{350} (\bibinfo{year}{2017}) \bibinfo{pages}{470--492}.
\bibitem[{Mendez et~al.(2018)Mendez, Ponga, and Ortiz}]{Mendez:2018}
\bibinfo{author}{J.~Mendez}, \bibinfo{author}{M.~Ponga},
  \bibinfo{author}{M.~Ortiz},
\newblock \bibinfo{title}{Diffusive molecular dynamics simulations of
  lithiation of silicon nanopillars},
\newblock \bibinfo{journal}{Journal of the Mechanics and Physics of Solids}
  \bibinfo{volume}{115} (\bibinfo{year}{2018}) \bibinfo{pages}{123--141}.
\bibitem[{Plimpton(1995)}]{LAMMPS}
\bibinfo{author}{S.~Plimpton},
\newblock \bibinfo{title}{Fast parallel algorithms for short-range molecular
  dynamics},
\newblock \bibinfo{journal}{Journal of Computational Physics}
  \bibinfo{volume}{117} (\bibinfo{year}{1995}) \bibinfo{pages}{1 -- 19}.
\bibitem[{Ponga and Sun(2018)}]{Ponga:2018}
\bibinfo{author}{M.~Ponga}, \bibinfo{author}{D.~Sun},
\newblock \bibinfo{title}{A unified framework for heat and mass transport at
  the atomic scale},
\newblock \bibinfo{journal}{Modelling and Simulation in Materials Science and
  Engineering} \bibinfo{volume}{26} (\bibinfo{year}{2018})
  \bibinfo{pages}{035014}.
\bibitem[{Martin(1990)}]{Martin:1990}
\bibinfo{author}{G.~Martin},
\newblock \bibinfo{title}{Atomic mobility in cahn's diffusion model},
\newblock \bibinfo{journal}{Phys. Rev. B} \bibinfo{volume}{41}
  (\bibinfo{year}{1990}) \bibinfo{pages}{2279--2283}.
\bibitem[{Zhang and Curtin(2008)}]{CurtinDiff:2008}
\bibinfo{author}{F.~Zhang}, \bibinfo{author}{W.~A. Curtin},
\newblock \bibinfo{title}{Atomistically informed solute drag in al-mg},
\newblock \bibinfo{journal}{Modelling and Simulation in Materials Science and
  Engineering} \bibinfo{volume}{16} (\bibinfo{year}{2008})
  \bibinfo{pages}{055006}.
\bibitem[{Kulkarni et~al.(2008)Kulkarni, Knap, and Ortiz}]{Kulkarni:2008}
\bibinfo{author}{Y.~Kulkarni}, \bibinfo{author}{J.~Knap},
  \bibinfo{author}{M.~Ortiz},
\newblock \bibinfo{title}{A variational approach to coarse graining of
  equilibrium and non-equilibrium atomistic description at finite temperature},
\newblock \bibinfo{journal}{Journal of the Mechanics and Physics of Solids}
  \bibinfo{volume}{56} (\bibinfo{year}{2008}) \bibinfo{pages}{1417 -- 1449}.
\bibitem[{Ariza et~al.(2012)Ariza, Romero, Ponga, and Ortiz}]{Ariza:2012}
\bibinfo{author}{M.~P. Ariza}, \bibinfo{author}{I.~Romero},
  \bibinfo{author}{M.~Ponga}, \bibinfo{author}{M.~Ortiz},
\newblock \bibinfo{title}{Hotqc simulation of nanovoid growth under tension in
  copper},
\newblock \bibinfo{journal}{International Journal of Fracture}
  \bibinfo{volume}{174} (\bibinfo{year}{2012}) \bibinfo{pages}{75--85}.
\bibitem[{Ponga et~al.(2012)Ponga, Romero, Ortiz, and Ariza}]{Ponga:2012}
\bibinfo{author}{M.~Ponga}, \bibinfo{author}{I.~Romero},
  \bibinfo{author}{M.~Ortiz}, \bibinfo{author}{M.~Ariza},
\newblock \bibinfo{title}{Finite temperature nanovoids evolution in fcc metals
  using quasicontinuum method},
\newblock in: \bibinfo{booktitle}{Key Engineering Materials}, volume
  \bibinfo{volume}{488}, \bibinfo{organization}{Trans Tech Publications}, pp.
  \bibinfo{pages}{387--390}.
\bibitem[{Ponga et~al.(2015)Ponga, Ortiz, and Ariza}]{Ponga:2015}
\bibinfo{author}{M.~Ponga}, \bibinfo{author}{M.~Ortiz},
  \bibinfo{author}{M.~Ariza},
\newblock \bibinfo{title}{Finite-temperature non-equilibrium quasi-continuum
  analysis of nanovoid growth in copper at low and high strain rates},
\newblock \bibinfo{journal}{Mechanics of Materials} \bibinfo{volume}{90}
  (\bibinfo{year}{2015}) \bibinfo{pages}{253--267}.
\bibitem[{Ponga et~al.(2016)Ponga, Ramabathiran, Bhattacharya, and
  Ortiz}]{Ponga:2016}
\bibinfo{author}{M.~Ponga}, \bibinfo{author}{A.~A. Ramabathiran},
  \bibinfo{author}{K.~Bhattacharya}, \bibinfo{author}{M.~Ortiz},
\newblock \bibinfo{title}{Dynamic behavior of nano-voids in magnesium under
  hydrostatic tensile stress},
\newblock \bibinfo{journal}{Modelling and Simulation in Materials Science and
  Engineering} \bibinfo{volume}{24} (\bibinfo{year}{2016})
  \bibinfo{pages}{065003}.
\bibitem[{Ponga et~al.(2017)Ponga, Ortiz, and Ariza}]{Ponga:2017}
\bibinfo{author}{M.~Ponga}, \bibinfo{author}{M.~Ortiz},
  \bibinfo{author}{M.~Ariza},
\newblock \bibinfo{title}{A comparative study of nanovoid growth in fcc
  metals},
\newblock \bibinfo{journal}{Philosophical Magazine}  (\bibinfo{year}{2017})
  \bibinfo{pages}{1--23}.
\bibitem[{Landau and Lifshitz(2013)}]{Landau:2013}
\bibinfo{author}{L.~Landau}, \bibinfo{author}{E.~Lifshitz},
  \bibinfo{title}{Statistical Physics}, \bibinfo{number}{v. 5},
  \bibinfo{publisher}{Elsevier Science}, \bibinfo{year}{2013}.
\bibitem[{Ullah and Ponga(2019)}]{Ullah:2019}
\bibinfo{author}{M.~W. Ullah}, \bibinfo{author}{M.~Ponga},
\newblock \bibinfo{title}{A new approach for electronic heat conduction in
  molecular dynamics simulations},
\newblock \bibinfo{journal}{Modelling and Simulation in Materials Science and
  Engineering} \bibinfo{volume}{27} (\bibinfo{year}{2019})
  \bibinfo{pages}{075008}.
\bibitem[{Ponga(2017)}]{webpage:maxent}
\bibinfo{author}{M.~Ponga}, \bibinfo{title}{{LAMMPS-MaxEnt} code description},
  \bibinfo{howpublished}{\url{http://mech-modsim.sites.olt.ubc.ca/software/ }},
  \bibinfo{year}{2017}. \bibinfo{note}{Accessed: 2017-08-31}.
\bibitem[{LAM(????)}]{LAMMPSDeveloperGuide}
\bibinfo{title}{Lammps developer guide},
  \bibinfo{howpublished}{\url{https://lammps.sandia.gov/doc/Developer.pdf}},
  ????
\bibitem[{Stukowski(2009)}]{Stukowski:2009}
\bibinfo{author}{A.~Stukowski},
\newblock \bibinfo{title}{Visualization and analysis of atomistic simulation
  data with {OVITO}{\textendash}the open visualization tool},
\newblock \bibinfo{journal}{Modelling and Simulation in Materials Science and
  Engineering} \bibinfo{volume}{18} (\bibinfo{year}{2009})
  \bibinfo{pages}{015012}.
\bibitem[{Crank(1979)}]{crank1979}
\bibinfo{author}{J.~Crank}, \bibinfo{title}{The mathematics of diffusion},
  \bibinfo{publisher}{Oxford university press}, \bibinfo{year}{1979}.
\bibitem[{Zhou et~al.(2004)Zhou, Johnson, and Wadley}]{Zhou:2004}
\bibinfo{author}{X.~W. Zhou}, \bibinfo{author}{R.~A. Johnson},
  \bibinfo{author}{H.~N.~G. Wadley},
\newblock \bibinfo{title}{Misfit-energy-increasing dislocations in
  vapor-deposited cofe/nife multilayers},
\newblock \bibinfo{journal}{Phys. Rev. B} \bibinfo{volume}{69}
  (\bibinfo{year}{2004}) \bibinfo{pages}{144113}.
\bibitem[{Gr\'egoire and Ponga(2017)}]{Gregoire:2017}
\bibinfo{author}{C.~Gr\'egoire}, \bibinfo{author}{M.~Ponga},
\newblock \bibinfo{title}{Nanovoid failure in magnesium under dynamic loads},
\newblock \bibinfo{journal}{Acta Materialia} \bibinfo{volume}{134}
  (\bibinfo{year}{2017}) \bibinfo{pages}{360 -- 374}.
\bibitem[{Mattsson et~al.(2009)Mattsson, Sandberg, Armiento, and
  Mattsson}]{Mattsson:2009}
\bibinfo{author}{T.~R. Mattsson}, \bibinfo{author}{N.~Sandberg},
  \bibinfo{author}{R.~Armiento}, \bibinfo{author}{A.~E. Mattsson},
\newblock \bibinfo{title}{Quantifying the anomalous self-diffusion in
  molybdenum with first-principles simulations},
\newblock \bibinfo{journal}{Phys. Rev. B} \bibinfo{volume}{80}
  (\bibinfo{year}{2009}) \bibinfo{pages}{224104}.
\bibitem[{Hirth and Lothe(1982)}]{Hirth}
\bibinfo{author}{J.~Hirth}, \bibinfo{author}{J.~Lothe}, \bibinfo{title}{Theory
  of Dislocations}, \bibinfo{publisher}{Krieger Publishing Company},
  \bibinfo{year}{1982}.
\bibitem[{Stukowski(2012)}]{CNA}
\bibinfo{author}{A.~Stukowski},
\newblock \bibinfo{title}{Structure identification methods for atomistic
  simulations of crystalline materials},
\newblock \bibinfo{journal}{Modelling and Simulation in Materials Science and
  Engineering} \bibinfo{volume}{20} (\bibinfo{year}{2012})
  \bibinfo{pages}{045021}.
\bibitem[{Stukowski et~al.(2012)Stukowski, Bulatov, and Arsenlis}]{DXA}
\bibinfo{author}{A.~Stukowski}, \bibinfo{author}{V.~Bulatov},
  \bibinfo{author}{A.~Arsenlis},
\newblock \bibinfo{title}{Automated identification and indexing of dislocations
  in crystal interfaces},
\newblock \bibinfo{journal}{Modelling and Simulation in Materials Science and
  Engineering} \bibinfo{volume}{20} (\bibinfo{year}{2012})
  \bibinfo{pages}{085007}.
\bibitem[{Liu et~al.(1997)Liu, Ohotnicky, Adams, Rohrer, and Hyland}]{Liu:1997}
\bibinfo{author}{X.-Y. Liu}, \bibinfo{author}{P.~Ohotnicky},
  \bibinfo{author}{J.~Adams}, \bibinfo{author}{C.~Rohrer},
  \bibinfo{author}{R.~Hyland},
\newblock \bibinfo{title}{Anisotropic surface segregation in alî—¸mg
  alloys},
\newblock \bibinfo{journal}{Surface Science} \bibinfo{volume}{373}
  (\bibinfo{year}{1997}) \bibinfo{pages}{357 -- 370}.
\bibitem[{Brandes(1992)}]{brandes:1992}
\bibinfo{author}{E.~Brandes}, \bibinfo{title}{G. b (editors), smithells metals
  reference book, butter worth}, \bibinfo{year}{1992}.
\bibitem[{Pellicer et~al.(2011)Pellicer, Varea, Sivaraman, Pan�, Suri�ach,
  Bar�, Nogu�s, Nelson, and Sort}]{Pellicer:2011}
\bibinfo{author}{E.~Pellicer}, \bibinfo{author}{A.~Varea},
  \bibinfo{author}{K.~M. Sivaraman}, \bibinfo{author}{S.~Pan�},
  \bibinfo{author}{S.~Suri�ach}, \bibinfo{author}{M.~D. Bar�},
  \bibinfo{author}{J.~Nogu�s}, \bibinfo{author}{B.~J. Nelson},
  \bibinfo{author}{J.~Sort},
\newblock \bibinfo{title}{Grain boundary segregation and interdiffusion effects
  in nickel-copper alloys: An effective means to improve the thermal stability
  of nanocrystalline nickel},
\newblock \bibinfo{journal}{ACS Applied Materials \& Interfaces}
  \bibinfo{volume}{3} (\bibinfo{year}{2011}) \bibinfo{pages}{2265--2274}.
  \bibinfo{note}{PMID: 21667966}.
\bibitem[{Sun et~al.(2017)Sun, Ariza, Ortiz, and Wang}]{sun2017acceleration}
\bibinfo{author}{X.~Sun}, \bibinfo{author}{M.~Ariza},
  \bibinfo{author}{M.~Ortiz}, \bibinfo{author}{K.~Wang},
\newblock \bibinfo{title}{Acceleration of diffusive molecular dynamics
  simulations through mean field approximation and subcycling time
  integration},
\newblock \bibinfo{journal}{Journal of Computational Physics}
  \bibinfo{volume}{350} (\bibinfo{year}{2017}) \bibinfo{pages}{470--492}.
\bibitem[{Sun et~al.(2019)Sun, Ariza, Ortiz, and Wang}]{sun2019atomistic}
\bibinfo{author}{X.~Sun}, \bibinfo{author}{M.~Ariza},
  \bibinfo{author}{M.~Ortiz}, \bibinfo{author}{K.~Wang},
\newblock \bibinfo{title}{Atomistic modeling and analysis of hydride phase
  transformation in palladium nanoparticles},
\newblock \bibinfo{journal}{Journal of the Mechanics and Physics of Solids}
  \bibinfo{volume}{125} (\bibinfo{year}{2019}) \bibinfo{pages}{360--383}.
\bibitem[{Sun et~al.(2018)Sun, Ariza, Ortiz, and Wang}]{sun2018long}
\bibinfo{author}{X.~Sun}, \bibinfo{author}{P.~Ariza},
  \bibinfo{author}{M.~Ortiz}, \bibinfo{author}{K.~G. Wang},
\newblock \bibinfo{title}{Long-term atomistic simulation of hydrogen absorption
  in palladium nanocubes using a diffusive molecular dynamics method},
\newblock \bibinfo{journal}{International Journal of Hydrogen Energy}
  \bibinfo{volume}{43} (\bibinfo{year}{2018}) \bibinfo{pages}{5657--5667}.
\bibitem[{Sun et~al.(2017)Sun, Ariza, Ortiz, and Wang}]{sun2017atomistic}
\bibinfo{author}{X.~Sun}, \bibinfo{author}{P.~Ariza},
  \bibinfo{author}{M.~Ortiz}, \bibinfo{author}{K.~G. Wang},
\newblock \bibinfo{title}{Atomistic simulation of hydrogen diffusion in
  palladium nanoparticles using a diffusive molecular dynamics method},
\newblock in: \bibinfo{booktitle}{ASME 2017 International Mechanical
  Engineering Congress and Exposition}, \bibinfo{organization}{American Society
  of Mechanical Engineers}, pp. \bibinfo{pages}{V009T12A026--V009T12A026}.
\bibitem[{Zhou et~al.(2008)Zhou, Zimmerman, Wong, and Hoyt}]{zhou2008}
\bibinfo{author}{X.~Zhou}, \bibinfo{author}{J.~Zimmerman},
  \bibinfo{author}{B.~Wong}, \bibinfo{author}{J.~Hoyt},
\newblock \bibinfo{title}{An embedded-atom method interatomic potential for
  pd–h alloys},
\newblock \bibinfo{journal}{Journal of Materials Research} \bibinfo{volume}{23}
  (\bibinfo{year}{2008}) \bibinfo{pages}{704–718}.
\bibitem[{Narayan et~al.(2017)Narayan, Hayee, Baldi, Koh, Sinclair, and
  A.~Dionne}]{Narayan2017}
\bibinfo{author}{T.~Narayan}, \bibinfo{author}{F.~Hayee},
  \bibinfo{author}{A.~Baldi}, \bibinfo{author}{A.~L. Koh},
  \bibinfo{author}{R.~Sinclair}, \bibinfo{author}{J.~A.~Dionne},
\newblock \bibinfo{title}{Direct visualization of hydrogen absorption dynamics
  in individual palladium nanoparticles},
\newblock \bibinfo{journal}{Nature Communications} \bibinfo{volume}{8}
  (\bibinfo{year}{2017}) \bibinfo{pages}{14020}.
\bibitem[{Stroud(1971)}]{Stroud:1971}
\bibinfo{author}{A.~Stroud}, \bibinfo{title}{Approximate calculation of
  multiple integrals}, Prentice-Hall series in automatic computation,
  \bibinfo{publisher}{Prentice-Hall}, \bibinfo{year}{1971}.
\bibitem[{Daw and Baskes(1984)}]{Daw:1984}
\bibinfo{author}{M.~S. Daw}, \bibinfo{author}{M.~I. Baskes},
\newblock \bibinfo{title}{Embedded-atom method: Derivation and application to
  impurities, surfaces, and other defects in metals},
\newblock \bibinfo{journal}{Phys. Rev. B} \bibinfo{volume}{29}
  (\bibinfo{year}{1984}) \bibinfo{pages}{6443--6453}.
\bibitem[{LeSar et~al.(1991)LeSar, Najafabadi, and Srolovitz}]{LeSar:1991}
\bibinfo{author}{R.~LeSar}, \bibinfo{author}{R.~Najafabadi},
  \bibinfo{author}{D.~J. Srolovitz},
\newblock \bibinfo{title}{Thermodynamics of solid and liquid embedded atom
  method metals: A variational study},
\newblock \bibinfo{journal}{The Journal of Chemical Physics}
  \bibinfo{volume}{94} (\bibinfo{year}{1991}) \bibinfo{pages}{5090--5097}.

\end{thebibliography}


\begin{thebibliography}{0}
\bibitem{ref1}  G. Venturini,  K. Wang,  I. Romero,  M. Ariza,  M. Ortiz,  Atomistic long-term simulation of heat and mass transport,  Journal of the Mechanics and Physics of Solids 73 (2014) 242 – 268.  
\bibitem{ref3} M. Ponga, D. Sun,  A unified framework for heat and mass transport at the atomic scale,  Modelling and Simulation in Materials Science and Engineering 26 (2018) 035014.        
\end{thebibliography}

\end{document}